\documentclass[12pt]{article}
\usepackage{amsfonts}
\usepackage{latexsym}
\usepackage{amsmath}
\usepackage{amssymb}
\usepackage{amssymb}

\newcommand{\newsection}{    
\setcounter{equation}{0}\section}
\def\appendix#1{\addtocounter{section}{1}\setcounter{equation}{0}
\renewcommand{\thesection}{\Alph{section}}
\section*{Appendix \thesection\protect\indent \parbox[t]{11.15cm}{#1}}
\addcontentsline{toc}{section}{Appendix \thesection\ \ \ #1}}

\newcommand{\be}{\begin{eqnarray}}
\newcommand{\ee}{\end{eqnarray}}
\newcommand{\bea}{\begin{eqnarray}}
\newcommand{\eea}{\end{eqnarray}}
\newcommand{\ba}{\begin{array}}
\newcommand{\ea}{\end{array}}
\newcommand{\nn}{\nonumber \\}

\newenvironment{frcseries}{\fontfamily{frc}\selectfont}{}

\def \la {\label}

\def\a{\alpha}
\def\b{\beta}
\def\l{\lambda}
\def\g{\gamma}

\def\e{\epsilon}

\def\cL{{\cal L}}

\def\bbe{{\bf{e}}}
\font\mybb=msbm10 at 11pt

\def\bb#1{\hbox{\mybb#1}}

\def\bZ {\bb{Z}}

\def\bC {\bb{C}}

\def\hn {{\tilde{\nabla}}}

\begin{document}
\begin{titlepage}
\begin{center}
\vspace*{-1.0cm}

\vspace{2.0cm} {\Large \bf  M-Horizons} \\[.2cm]

\vspace{1.5cm}
 {\large  J. Gutowski and  G. Papadopoulos}

\vspace{0.5cm}
Department of Mathematics\\
King's College London\\
Strand\\
London WC2R 2LS, UK\\

\vspace{0.5cm}

\end{center}

\vskip 1.5 cm
\begin{abstract}
\end{abstract}
We solve the Killing spinor equations and determine the near horizon geometries of M-theory that preserve at least one supersymmetry.  The M-horizon spatial sections are 9-dimensional manifolds
with a $Spin(7)$ structure restricted by  geometric constraints which we give explicitly.  We also provide an alternative characterization of the solutions of the Killing spinor equation, utilizing the compactness of the horizon section and the field equations,  by proving a Lichnerowicz type of
theorem which implies that the zero modes of a Dirac operator coupled to 4-form fluxes are Killing spinors.
We use  this, and  the maximum  principle,  to  solve the field equations of the theory for some special cases and present some examples.
\end{titlepage}


\section{Introduction}

Most of the supergravity solutions that have widespread applications in strings and M-theory, such as branes and black holes, admit stationary time-like  Killing vector fields which become null at a hypersurface, ie exhibit Killing horizons. For extreme branes \cite{gibbons} and black holes, a suitable geometry can be defined near the Killing horizons, ``the near horizon geometry,''
that is also a solution of the supergravity field equations. Such near horizon geometries  exhibit additional symmetries to those of the original brane and black hole spacetimes
 and they have been the focus of intense investigations in the context of AdS/CFT \cite{maldacena} and black hole thermodynamics.

 Under certain regularity assumptions,  one can adapt Eddington-Finkelstein type of coordinates  near every Killing horizon \cite{isen}, \cite{gnull}. In these coordinates and  in the near horizon limit,  the metric and fluxes of brane and black hole solutions   take a particularly simple form. As a result, it is more straightforward to construct the near horizon geometries than those of brane and black hole spacetimes. Because of this, the near horizon geometries can also be used to provide evidence for the existence of new
 solutions and explore uniqueness  theorems for black hole and brane solutions in diverse dimensions \cite{israel}-\cite{obers1}.

The search for near horizon geometries in supergravity theories has been facilitated by the additional assumption that they preserve at least one supersymmetry. Using
this, a systematic investigation of such solutions can be made, following the solution of the problem for the simple 5-dimensional supergravity in \cite{reallbh} and the subsequent construction
of non-spherical black hole solutions, the black rings in \cite{ring}.
The topology,  geometry and fractions of supersymmetry preserved by the horizons of ${\cal N}=1$ $d=4$, $(1,0)$ $d=6$ and  heterotic supergravities  have been determined \cite{d4hor, d6hor, hethor}, respectively.
Moreover those horizons that preserve at least half of the supersymmetry have been classified up to a local  isometry. These results have followed the solution
of the Killing spinor equations (KSEs) of these theories for backgrounds preserving any number of supersymmetries \cite{clasd4, ortin, clasd6, clashet}. In addition, the topology and geometry of IIB horizons with 5-form flux has been identified in \cite{iibhor}
and that of static M-horizons in \cite{smhor}.  In these two cases, it has been assumed that the solutions preserve at least one supersymmetry. Again the solution of the KSEs of IIB \cite{iibclas} and 11-d supergravity \cite{pakis, ggp} for backgrounds preserving one supersymmetry  has been utilized. However all our calculations
rely on the spinorial geometry technique of \cite{ggp}. It is not known how to identify
all the fractions  of supersymmetry preserved by IIB and M-horizons.

In this paper, we solve the KSEs of M-horizons preserving at least one supersymmetry extending the results for static horizons in \cite{smhor}. For this, we identify the black hole stationary Killing vector field
with the Killing vector field of backgrounds preserving one supersymmetry constructed as a Killing spinor bilinear.
We shall find that the horizon spatial sections ${\cal S}$ admit a $Spin(7)$ structure which satisfies some geometric constraints  given in either  (\ref{q12}) or (\ref{q12b}). Moreover we show
some of the fluxes are expressed in terms of the geometry but some others remain unrestricted by the KSEs. The full solution of the KSEs can be found in appendix A and it is expressed in a gauge where
there is a preferred direction\footnote{Such a direction is always attained as ${\cal S}$ is
 9-dimensional, the Euler number vanishes, and so it admits  an everywhere non-vanishing vector field.} in the tangent space of ${\cal S}$.

We also give a covariant description of the  solution to the KSE by proving a Lichnerowicz type of theorem.  The KSE of 11-dimensional supergravity is a parallel transport equation with respect to the supercovariant connection
${\cal D}$ associated with the gravitino supersymmetry transformation.  It is apparent that every solution of the KSE is also a solution of the  Dirac equation
associated with ${\cal D}$ which exhibits a coupling to the 4-form field strength. Using the compactness of the spatial horizon sections and the field equations of the theory, we shall demonstrate that every zero mode of the Dirac equation
on the horizon section gives rise to a parallel and so Killing spinor.

To find examples of near horizon geometries, one has to also solve the field equations and Bianchi identities of the theory which are not implied as integrability conditions of KSEs. In many theories of interest, like heterotic,  the former are solved or significantly simplified by using the maximum principle utilizing a scalar on the spatial horizon sections that is constructed from the data of the problem. It is not apparent that all M-horizon geometries  can be found in this way. Nevertheless, we explore some special cases relying on the maximum principle to solve or simplify the field equations. In particular, we focus on magnetic horizons and give some  explicit results. In addition, we show
that all heterotic horizons can be lifted to M-horizons. This class of M-horizons exhibits a supersymmetry enhancement, ie it preserves at least 2 supersymmetries,  which is  a consequence of a similar property  for heterotic horizons.

This paper is organized as follows. In section two, we describe the near horizon fields, state our notation and give the field equations of 11-d supergravity. In section 3, we solve the KSEs. Part of  the
results of this section are summarized in appendix A.  In section 4, we prove the Lichnerowicz type of
theorem. Again some of the computations for this are presented in appendices B and C. In section 5,
we explore the magnetic horizons and give some examples. In section 6, we also describe the lifting
of heterotic horizons to M-theory horizons, and in section 7 we give our conclusions. In appendix
D, we solve the KSEs of an example presented in section 5.

\section{Near Horizon Geometry}

\subsection{Near horizon fields}

A straightforward adaptation of the analysis in \cite{isen}, \cite{gnull} gives that the  metric of the near M-horizon geometries
 can be written as
 \bea
ds^2 &=& 2 \bbe^+ \bbe^- + \delta_{ij} \bbe^i \bbe^j=2 du (dr + r h - {1 \over 2} r^2 \Delta du)+ g_{IJ} dy^Idy^J~,
\la{mhm}
\eea
where we have introduced the frame
\bea
\bbe^+ = du~,~~~\bbe^- = dr + r h - {1 \over 2} r^2 \Delta du~,~~~
\bbe^i = e^i{}_J dy^J~;~~~g_{IJ}=\delta_{ij} e^i{}_I e^j{}_J~,
\la{nhbasis}
\eea
 the dependence on the coordinates $r,u$ is given explicitly, and $h=h_i \bbe^i, \Delta$ and $e^i_I$ depend only on the rest of the coordinates $y$.
We choose the frame indices  $i=1, 2, 3, 4, 6, 7, 8, 9, \sharp$ and we follow the conventions of \cite{system11}.
Observe that the Killing vector field $\partial_u$ is non-space-like everywhere as $\Delta\geq 0$,  and becomes null at $r=0$. There is no loss of generality
in taking $\Delta\geq 0$ as this is implied by the KSEs. The near horizon section ${\cal S}$ is the co-dimension 2 subspace given by $r=u=0$
equipped with the metric
\bea
ds^2({\cal S})=g_{IJ} dy^Idy^J~,
\eea
and it is assumed to be compact, connected and without boundary. The near horizon metric (\ref{mhm}) has an additional isometry
to $\partial_u$, associated with the scaling transformation $r \rightarrow \ell r$  $u \rightarrow \ell^{-1} u$, which may not be extended beyond the near horizon limit.

Now let us turn to the 3-form gauge potential $C$ of the M-horizons. Using the extremality condition  $C_{-ij}=0$ and that the field strength $F=dC$ must must be invariant
under both the stationary and scaling isometries of the near horizon geometries, the most general form of $C$ is
\be
C = r \bbe^+ \wedge B + \bbe^+ \wedge \bbe^- \wedge A + G~,
\ee
where $A$, $B$, $G$ are $u, r$-independent 1-, 2- and 3-forms on ${\cal{S}}$, respectively.
Setting\footnote{If $L$ is a k-form, then $d_hL=dL-h\wedge L$.}
\be
Y = -B  + d_hA~,
\ee
one obtains
\be
F = \bbe^+ \wedge \bbe^- \wedge Y
+ r \bbe^+ \wedge d_h Y + X~,
\la{mhf}
\ee
where $Y \in \Lambda^2({\cal{S}})$, $X \in \Lambda^4 ({\cal{S}})$ are $u, r$-independent
two and four forms,  respectively.
The remaining condition imposed by the Bianchi identity is
\bea
\label{clos}
dX=0 \ .
\eea
To summarize, the metric and 4-form field strength of M-horizons can be expressed as in (\ref{mhm}) and (\ref{mhf}), respectively.

\subsection{Field Equations}

The field equations of 11-dimensional supergravity \cite{julia} for M-horizons decompose along the light-cone  and spatial horizon section ${\cal S}$ directions.
The field equation of the 3-form gauge potential is
\be
d \star_{11}  F -{1 \over 2} F \wedge F=0~,
\ee
where $\star_{11}$ is the Hodge star operation of 11-dimensional spacetime.
These can be decomposed as
\be
\label{geq1}
-\star_9 d_hY  - h \wedge \star_9 X + d \star_9 X = Y \wedge X~,
\ee
and
\be
\label{geq2}
-d \star_{9} Y = {1 \over 2} X \wedge X~,
\ee
where $\star_{9}$ is the Hodge star operation on ${\cal S}$,  spacetime volume form  is chosen as $\epsilon_{11}= \bbe^+ \wedge \bbe^- \wedge \epsilon_{{\cal{S}}}$  and $\epsilon_{{\cal{S}}}$
is the volume form of ${\cal S}$. Equivalently, in components, one has
\bea
\hn^i X_{i \ell_1 \ell_2 \ell_3} + 3 \hn_{[\ell_1} Y_{\ell_2 \ell_3]}
=3 h_{[\ell_1} Y_{\ell_2 \ell_3]} + h^i X_{i \ell_1 \ell_2 \ell_3} -
{1 \over 48} \epsilon_{\ell_1 \ell_2 \ell_3}{}^{q_1 q_2 q_3 q_4 q_5 q_6}
Y_{q_1 q_2} X_{q_3 q_4 q_5 q_6}~,
\eea
and
\bea
\hn^j Y_{ji}-{1\over 1152} \epsilon_{i}{}^{q_1 q_2 q_3 q_4 q_5 q_6 q_7 q_8} X_{q_1 q_2 q_3 q_4} X_{q_5 q_6 q_7 q_8}=0~,
\eea
where $\hn$ is the Levi-Civita connection of the metric on the near horizon section ${\cal{S}}$.

The Einstein equation is
\be
R_{MN} = {1 \over 12} F_{M L_1 L_2 L_3} F_N{}^{L_1 L_2 L_3}
-{1 \over 144} g_{MN} F_{L_1 L_2 L_3 L_4}F^{L_1 L_2 L_3 L_4} \ .
\ee
This decomposes into a number of components. In particular along ${\cal{S}}$, one finds
\bea
\label{ein1}
{\tilde {R}}_{ij} + \hn_{(i} h_{j)} -{1 \over 2} h_i h_j &=& -{1 \over 2} Y_{i \ell} Y_j{}^\ell
+{1 \over 12} X_{i \ell_1 \ell_2 \ell_3} X_j{}^{\ell_1 \ell_2 \ell_3}
\nonumber \\
&+& \delta_{ij} \bigg( {1 \over 12} Y_{\ell_1 \ell_2} Y^{\ell_1 \ell_2}
-{1 \over 144} X_{\ell_1 \ell_2 \ell_3 \ell_4} X^{\ell_1 \ell_2 \ell_3 \ell_4} \bigg)~,
\eea
where  ${\tilde{R}}_{ij}$ is the Ricci tensor of ${\cal{S}}$.
The $+-$ component of the Einstein equation gives
\bea
\label{einpm}
\hn^i h_i = 2 \Delta + h^2 -{1 \over 3} Y_{\ell_1 \ell_2} Y^{\ell_1 \ell_2}
-{1 \over 72} X_{\ell_1 \ell_2 \ell_3 \ell_4} X^{\ell_1 \ell_2 \ell_3 \ell_4}~.
\eea
Similarly, the $++$ and $+i$ components of the Einstein equation can be expressed as
\bea
\label{einpp}
{1 \over 2} \hn^i \hn_i \Delta -{3 \over 2} h^i \hn_i \Delta -{1 \over 2} \Delta
\hn^i h_i + \Delta h^2 +{1 \over 4} dh_{ij} dh^{ij}
= {1 \over 12} (d_hY)_{\ell_1 \ell_2 \ell_3} (d_h Y)^{\ell_1 \ell_2 \ell_3}~,
\nonumber \\
\eea
and
\bea
\label{einpi}
-{1 \over 2} \hn^j dh_{ji} + h^j (dh)_{ji} - \hn_i \Delta + \Delta h_i
= {1 \over 12} X_i{}^{\ell_1 \ell_2 \ell_3} (d_h Y)_{\ell_1 \ell_2 \ell_3}
-{1 \over 4} (d_h Y)_i{}^{\ell_1 \ell_2} Y_{\ell_1 \ell_2} \ ,
\nonumber \\
\eea
respectively.
Although we have included the $++$ and the $+i$ components of the
Einstein equations for completeness, it is straightforward to show that
both ({\ref{einpp}}) and ({\ref{einpi}}) hold as a consequence of
({\ref{clos}}), the 3-form field equations ({\ref{geq1}}) and ({\ref{geq2}}) and
the components of the Einstein equation in ({\ref{ein1}}) and  ({\ref{einpm}}).
This does not make use of supersymmetry, or any assumptions on the topology of
${\cal{S}}$.
Hence, the conditions on $\Delta$, $h$, $Y$ and $X$ simplify to
({\ref{clos}}), ({\ref{geq1}}), ({\ref{geq2}}), ({\ref{ein1}}) and ({\ref{einpm}}).

\newsection{Killing spinor equations}

The KSE of 11-dimensional supergravity is
\bea
\nabla_M \epsilon
+\bigg(-{1 \over 288} \Gamma_M{}^{L_1 L_2 L_3 L_4} F_{L_1 L_2 L_3 L_4}
+{1 \over 36} F_{M L_1 L_2 L_3} \Gamma^{L_1 L_2 L_3} \bigg) \epsilon =0~,
\nonumber \\
\eea
where $\nabla$ is the spacetime Levi-Civita connection.  The KSE can be decomposed along the
light-cone and ${\cal S}$ directions, as in the decomposition of the field equations.
Before we proceed with the analysis,
the non-vanishing components of the spin connection for the metric (\ref{mhm})  are
\bea
\Omega_{+,+-} &=& -r \Delta, \quad \Omega_{+,+i} = {1 \over 2} r^2 (\Delta h_i - \partial_i \Delta),
\quad \Omega_{+,-i} = -{1 \over 2} h_i, \quad \Omega_{+,ij} = -{1 \over 2} r (dh)_{ij}~,
\cr
\Omega_{-,+i} &=& -{1 \over 2} h_i~,~~~
\Omega_{i,+-} ={1 \over 2} h_i, \quad \Omega_{i,+j} = -{1 \over 2} r (dh)_{ij},
\qquad \Omega_{i,jk} = \tilde\Omega_{i,jk}~,
\eea
where $\tilde\Omega_{i,jk}$ is the spin connection of the horizon section  ${\cal S}$.

To solve the KSEs for M-horizons,  we shall first demonstrate that they can be integrated along the light-cone directions. Then, we shall assume
that the stationary Killing vector field $\partial_u$ is identified with the vector Killing spinor bi-linear of backgrounds
preserving one supersymmetry. This leads to a simplification of the KSEs along the horizon section ${\cal S}$ which we solve using  spinorial geometry \cite{ggp}.
For the analysis, we use the spinor conventions and the ``null'' spinor basis of Appendix A in  \cite{system11}.

\subsection{Integrability of light-cone directions}

To solve the KSEs along the light-cone directions, we set
\be
\epsilon = \epsilon_+ + \epsilon_-~,~~~\Gamma_\pm \epsilon_\pm =0~.
\ee
Then after some computation, we find that
\bea
\label{ksp1}
\epsilon_+ = \eta_+, \qquad \epsilon_- = \eta_- + r \Gamma_-
\Theta_+\eta_+~,
\eea
and
\bea
\label{ksp2}
\eta_+ = \phi_+ + u \Gamma_+ \Theta_- \phi_- , \qquad \eta_- = \phi_-~,
\eea
where
\bea
\Theta_\pm=\bigg({1 \over 4} h_i \Gamma^i +{1 \over 288} X_{\ell_1 \ell_2 \ell_3 \ell_4}
\Gamma^{\ell_1 \ell_2 \ell_3 \ell_4} \pm {1 \over 12} Y_{\ell_1 \ell_2} \Gamma^{\ell_1 \ell_2} \bigg)~,
\eea
and $\phi_\pm = \phi_\pm (y)$ do not depend on $r$ or $u$.

Furthermore, the $+$ and $-$
components of the KSE impose the following algebraic conditions on the Killing spinors
\bea
\label{cc1}
&&\bigg( {1 \over 2} \Delta -{1 \over 8} dh_{ij} \Gamma^{ij} +{1 \over 72} d_hY_{\ell_1 \ell_2 \ell_3}
\Gamma^{\ell_1 \ell_2 \ell_3}
\nn
&+&2 \big({1 \over 4} h_i \Gamma^i -{1 \over 288} X_{\ell_1 \ell_2 \ell_3 \ell_4}
\Gamma^{\ell_1 \ell_2 \ell_3 \ell_4} +{1 \over 12} Y_{\ell_1 \ell_2} \Gamma^{\ell_1 \ell_2} \big)
\Theta_+ \bigg) \phi_+ =0~,
\eea

\bea
\label{cc2}
&& \bigg({1 \over 4} \Delta h_i \Gamma^i - {1 \over 4} \partial_i \Delta \Gamma^i
+ \big( -{1 \over 8} dh_{ij} \Gamma^{ij} -{1 \over 24} d_hY_{\ell_1 \ell_2 \ell_3} \Gamma^{\ell_1 \ell_2
\ell_3} \big)\,
\Theta_+ \bigg) \phi_+ =0~,
\eea

\bea
\label{cc3}
&&\bigg( {1 \over 2} \Delta -{1 \over 8} dh_{ij} \Gamma^{ij} -{1 \over 72} d_hY_{\ell_1 \ell_2 \ell_3}
\Gamma^{\ell_1 \ell_2 \ell_3}
\nn
&-&2 \Theta_-
\big(-{1 \over 4} h_j \Gamma^j +{1 \over 288} X_{n_1 n_2 n_3 n_4}
\Gamma^{n_1 n_2 n_3 n_4} +{1 \over 12} Y_{n_1 n_2} \Gamma^{n_1 n_2} \big) \bigg)
\Theta_-
\phi_- =0~,
 \eea

\bea
\label{cc4}
&& \bigg(-{1 \over 4} \Delta h_i \Gamma^i + {1 \over 4} \partial_i \Delta \Gamma^i
+ \big( -{1 \over 8} dh_{ij} \Gamma^{ij} +{1 \over 24} d_hY_{\ell_1 \ell_2 \ell_3} \Gamma^{\ell_1 \ell_2
\ell_3} \big)
\nn
&& \big(-{1 \over 4} h_n \Gamma^n +{1 \over 288} X_{n_1 n_2 n_3 n_4}
\Gamma^{n_1 n_2 n_3 n_4} +{1 \over 12} Y_{n_1 n_2} \Gamma^{n_1 n_2} \big) \bigg)
\Theta_-
\phi_- =0~,
 \eea

\bea
\label{cc5}
&& \bigg( -{1 \over 2} \Delta -{1 \over 8} dh_{ij} \Gamma^{ij} +{1 \over 24}
d_hY_{\ell_1 \ell_2 \ell_3} \Gamma^{\ell_1 \ell_2 \ell_3}
\nn
&+& 2 \big(-{1 \over 4} h_n \Gamma^n +{1 \over 288} X_{n_1 n_2 n_3 n_4}
\Gamma^{n_1 n_2 n_3 n_4} +{1 \over 12} Y_{n_1 n_2} \Gamma^{n_1 n_2} \big)
\Theta_- \bigg)
\phi_- =0~.
 \eea
Since we have separated the light-cone directions from the rest, the remaining KSEs have manifest $Spin(9)\subset Spin(10,1)$ local gauge
invariance.

\subsection{Horizons with one supersymmetry}

\subsubsection{Stationary Killing vector field and spinor bilinears}\label{VeW}

The associated 1-form of the stationary Killing vector field $\partial_u$ of M-horizon geometries is
\bea
V=\bbe^--{1 \over 2} r^2 \Delta \bbe^+~,
\eea
as expressed in the basis ({\ref{nhbasis}}). For backgrounds preserving at least one supersymmetry,
the spacetime 1-form constructed as a Killing spinor bilinear is identified with
\be
W= \langle B \epsilon^*, \Gamma_A \epsilon \rangle\,\bbe^A \ .
\ee

For supersymmetric black holes\footnote{The identification of $V$ with $W$ is up to a constant scale chosen at convenience.}, $V=W$. To find the conditions implied by this identification, we shall use the local $Spin(9)$ gauge invariance of the
KSE, after solving along the light-cone directions, to choose $\phi_\pm$ and identify $W$. For this
observe that both $\phi_\pm$ are Majorana $Spin(9)$ spinors. Moreover it is well known that $Spin(9)$ acts
transitively on the $S^{15}$ sphere in the Majorana representation and the isotropy group is $Spin(7)$, ie $Spin(9)/Spin(7)=S^{15}$.
Using this, one can orient the spinor $\phi_-$ along any direction with a $Spin(9)$ transformation. As a result, we can choose
\be
\phi_- = w (e_5+e_{12345})~,
\ee
where $w$ is a real function.  Next, on comparing the components of $W$ with those of $V$ in the basis ({\ref{nhbasis}}),
we require that
\be
W_+|_{r=0} =0~,
\ee
which in turn imposes the condition $w=0$. So we conclude that
\be
\phi_-=0~.
\ee

As $\phi_-=0$ is invariant under $Spin(9)$, the $Spin(9)$ gauge transformations can be used again to choose $\phi_+$. In particular, without loss of generality,
one can set
\be
\phi_+ = g(1+e_{1234})~,
\ee
for some real $r,u$-independent function $g$.  Note then that
\be
W_- = -2 \sqrt{2} g^2~.
\ee
As we require that $V_-=1$, this implies that $g$ is constant. For convenience, we set $g=1$, so that
\be
\phi_+ = 1+e_{1234}~.
\ee

To proceed, note that as a consequence of ({\ref{ksp1}}) and ({\ref{ksp2}}), it follows that
\be
\label{kspin}
\label{thetdef}
\epsilon = \phi_+ +r \Gamma_- \zeta~,~~~\zeta \equiv  \Theta_+\phi_+~.
\ee

Next, the condition $W_i=0$ implies that
\bea
\langle \phi_+ , \Gamma_i \zeta \rangle =0~,
\la{wio}
\eea
or equivalently
\bea
\label{blmatch1}
h_i \langle \phi_+ , \phi_+ \rangle +{2 \over 3} Y_{i \ell} \langle \phi_+ , \Gamma^\ell  \phi_+ \rangle +{1 \over 72} \langle \phi_+ , \Gamma_{i}{}^{\ell_1 \ell_2 \ell_3 \ell_4}
X_{\ell_1 \ell_2 \ell_3 \ell_4} \phi_+ \rangle =0 \ .
\eea
The remaining condition imposed by comparing $V$ and $W$ is obtained from $V_+=W_+$ as
\bea
\label{blmatch3}
{1 \over 2} \langle \phi_+ , \phi_+ \rangle \Delta = 2 \langle \zeta , \zeta \rangle~.
\eea
This concludes the analysis of the conditions which arise from the identification of the stationary Killing vector field with that constructed
as a Killing spinor bilinear.

\subsubsection{KSEs along ${\cal S}$}

The  KSEs along the spatial horizon section ${\cal S}$ can be written as
\bea
\label{sp1}
\hn_i \phi_+ + \bigg( -{1 \over 4} h_i -{1 \over 288} \Gamma_i{}^{\ell_1 \ell_2 \ell_3 \ell_4}
X_{\ell_1 \ell_2 \ell_3 \ell_4} +{1 \over 36} X_{i \ell_1 \ell_2 \ell_3} \Gamma^{\ell_1 \ell_2 \ell_3}
\nn
+{1 \over 24} \Gamma_i{}^{\ell_1 \ell_2} Y_{\ell_1 \ell_2} -{1 \over 6} Y_{ij}\Gamma^j \bigg) \phi_+ =0~,
\eea
and
\bea
\label{sp2}
\hn_i \zeta +  \bigg(-{1 \over 2} h_i +{1 \over 4} \Gamma_i{}^\ell h_\ell
-{1 \over 24} X_{i \ell_1 \ell_2 \ell_3} \Gamma^{\ell_1 \ell_2 \ell_3}
+{1 \over 8} \Gamma_i{}^{\ell_1 \ell_2} Y_{\ell_1 \ell_2} \bigg) \zeta
\nn
+ \bigg( {1 \over 4} \Delta \Gamma_i -{1 \over 16} \Gamma_i{}^{\ell_1 \ell_2} dh_{\ell_1 \ell_2}
-{3 \over 8} dh_{i \ell}\Gamma^{\ell} -{1 \over 48} d_hY_{\ell_1 \ell_2 \ell_3}
\Gamma^{\ell_1 \ell_2 \ell_3} \Gamma_i \bigg) \phi_+ =0~.
\nonumber \\
\eea
As we shall show (\ref{sp2}) is not an independent condition.

\subsection{Solution of KSEs}

\subsubsection{Preliminaries}

Before we proceed to describe the solution to the KSEs, it is instructive to specify the independent ones. Since $\phi_-=0$, the conditions (\ref{cc3}) to (\ref{cc5})
are automatically satisfied. Moreover, the conditions ({\ref{cc1}}) and ({\ref{cc2}}) are implied by ({\ref{sp1}}) and ({\ref{sp2}}) as well as the conditions obtained
from the identification $V=W$ in section \ref{VeW}. In particular, the elimination of ({\ref{cc1}}) and ({\ref{cc2}}) as independent conditions
 follows without the use of field equations and Bianchi identities. The details of this analysis are presented in Appendix A.

Further simplification is possible provided that the field equations and the Bianchi identities are used as well. In particular,  one can  show that ({\ref{sp2}}) is implied
by ({\ref{sp1}}),  after using the Bianchi identity $dX=0$ ({\ref{clos}}), the 3-form gauge potential field equations
({\ref{geq1}}) and ({\ref{geq2}}), and the components of the Einstein equations on ${\cal{S}}$.
The details of this are presented in Appendix B.

Furthermore, the $+-$ component of the Einstein equation ({\ref{einpm}}) is implied by
supersymmetry.  This component has not been used to rewrite any of the conditions involving
the Killing spinor. It is straightforward to show that ({\ref{einpm}}) is obtained by contracting
({\ref{redx1}}) with $\Gamma^i$, and then using ({\ref{clos}}), ({\ref{geq1}}) and ({\ref{geq2}}).

Next, observe that if $\phi=\phi_+$ satisfies ({\ref{sp1}}), then it follows that
\bea
\label{constnorm1}
\hn_i \langle \phi , \phi \rangle = {1 \over 2} h_i \langle \phi , \phi \rangle
+{1 \over 3} Y_{i \ell} \langle \phi, \Gamma^\ell \phi \rangle +{1 \over 144}
\langle \phi , \Gamma_i{}^{\ell_1 \ell_2 \ell_3 \ell_4} X_{\ell_1 \ell_2 \ell_3 \ell_4} \phi \rangle~.
\eea
Hence $\langle \phi , \phi \rangle$ is constant iff
\bea
h_i \langle \phi , \phi \rangle +{2 \over 3} Y_{i \ell} \langle \phi , \Gamma^\ell  \phi \rangle +{1 \over 72} \langle \phi , \Gamma_{i}{}^{\ell_1 \ell_2 \ell_3 \ell_4}
X_{\ell_1 \ell_2 \ell_3 \ell_4} \phi \rangle =0 \ .
\eea
This condition is identical to ({\ref{blmatch1}}) which has been derived from the identification $V=W$ in section \ref{VeW}.
Hence it follows that provided the Bianchi identity ({\ref{clos}})
and bosonic field equations ({\ref{geq1}}), ({\ref{geq2}}) and ({\ref{ein1}}) hold, the conditions imposed by supersymmetry reduce to ({\ref{sp1}}) and ({\ref{blmatch3}}), where $\zeta$ and the Killing spinor $\e$ are  given in ({\ref{thetdef}}).

We remark that the condition $\langle \phi, \phi \rangle =\mathrm {const}$, which as we have previously observed arises from
the identification of the Killing vectors $V=W$ in section \ref{VeW},  can also be derived. In particular
one can take the divergence of  ({\ref{constnorm1}}) and then  use  ({\ref{sp1}}),
({\ref{einpm}}) to express $\hn^i h_i$ in terms of $h$, $Y$ and $X$, and  ({\ref{geq2}}) to rewrite
$\hn^i Y_{ij}$ in terms of $X \wedge X$. One then obtains the condition
\bea
\hn^i \hn_i \langle \phi, \phi \rangle -2 h^i \hn_i \langle \phi, \phi \rangle =0 \ .
\eea
On applying the maximum principle, it follows that  $\langle \phi, \phi \rangle = \mathrm{const}$.

\subsubsection{Solution of the linear system}
As we have mentioned there are several ways to choose the independent KSEs to solve. Perhaps the most straightforward case is to consider the KSEs ({\ref{sp1}}) and
({\ref{sp2}}) together with $\phi_+=1+e_{1234}$, which arises from the identification $V=W$ in section \ref{VeW},  and to use the remaining $Spin(7)$ gauge invariance to choose $\zeta$ in (\ref{thetdef}).
 This can be done as follows. The identification $W=V$ implies that $W_i=0$, which in turn can be written as (\ref{wio}).  This implies that
$\zeta$ must be a linear combination of $i(1-e_{1234})$ and $e_{ij}$, ie $\zeta$ lies in the ${\bf 7}$
representation of $Spin(7)$. In such a case, a $Spin(7)$ transformation can be used, which leaves $1+e_{1234}$ invariant, such that without loss of generality
one can set
\be
\label{gfix}
 \zeta= \Theta_+ \phi_+ = i \Phi (1-e_{1234})~,
\ee
for some  $\Phi$ a real function on ${\cal S}$.  Thus the Killing spinor is
\bea
\e=1+e_{1234}+ i r \Phi (1-e_{1234})~.
\la{gfixks}
\eea

Using the above gauge, the KSEs ({\ref{sp1}}) and ({\ref{sp2}}) can be expressed as a linear system which has been presented in appendix A.
The linear system can be solved to express some of the fluxes in terms of the geometry and find the conditions on the geometry of ${\cal S}$ imposed by supersymmetry.
There are two cases to consider. First suppose that as some patch $\Delta\not=0$. In such a case, one has from (\ref{blmatch3}) that
\bea
\Delta=4\Phi^2~.
\eea
The solution of the linear system is explicitly given in appendix A. The solution of the system for $\Delta=0$ is also given in appendix A.
In both cases, not all fluxes of the theory are determined in terms of the geometry. In particular,  the (2,2) and traceless part of $X$
is not determined by the KSEs.

\subsubsection{Topology and Geometry of horizon sections}

The spatial horizon section ${\cal S}$ admits a $Spin(7)$ structure. This is because the parallel transport equation (\ref{sp1}) implies that there is
a nowhere vanishing spinor $\phi_+$ on ${\cal S}$. However, this does  not impose a topological condition on ${\cal S}$ as every 9-dimensional spin manifold
admits a $Spin(7)$ structure. To see this, the structure group of ${\cal S}$ will reduce to $Spin(7)$ iff the
 the  bundle $P\times_{Spin(9)}  Spin(9)/Spin(7)$ with fibre $Spin(9)/Spin(7)$
 admits a section, where $P$ is the principal spin bundle over ${\cal S}$. The principal topological obstructions for the existence of
 such sections lie in $H^k({\cal S}, \pi_{k-1}(Spin(9)/Spin(7)))$. But $Spin(9)/Spin(7)=S^{15}$ and so $\pi_{k}(Spin(9)/Spin(7))=0$ for $k<15$.
 As a result all obstructions vanish.

Observe also that if $\Delta\not=0$ everywhere on ${\cal S}$, then there is an additional nowhere vanishing spinor on ${\cal S}$ and its structure group reduces
to $SU(4)$. Unlike the previous case, there are topological obstructions to reducing the structure group of ${\cal S}$ to $SU(4)$ the first lying in $H^6({\cal S}, \bZ)$.
But  $\Delta\not=0$ is an additional condition and it does not always follow  from the KSEs.

The only geometric condition on ${\cal S}$ linear in the spin connection which follows from the solution of the linear system is (\ref{q12}).
As  has been mentioned in appendix A, there are additional conditions on the geometry which however are quadratic in the spin connection
and as a result resemble integrability conditions.

Although the linear system associated with the KSEs can be solved and the geometric conditions on the horizon section ${\cal S}$ can be identified,
the rather involved form of the resulting equations which express the fluxes in terms of the geometry  do not yield a manageable expression after  substitution into the field equations\footnote{The same applies when one attempts to use the solutions of the KSEs to apply  the maximum
principle  on either  $\Delta$ or $h^2$. In any case, it is not apparent that all M-horizons
 have either $\Delta$ or $h^2$ constant that an application of the maximum principle would imply.
 In fact, we know that there are static horizons \cite{smhor} with $h^2$ non-constant given by the warped
 $AdS_2$ solutions of \cite{kim}.}. In addition, the choice of gauge we have made leads to a distinguished direction $\bbe^\sharp$ on ${\cal S}$.
 This works well for the M-horizons which manifestly exhibit such a direction\footnote{Topologically, there is always such
 a direction on a 9-dimensional manifold as its Euler number vanishes. However, geometrically
 such a direction may not be manifest.}, like those for example that are associated
  with either IIA or heterotic horizons.  Otherwise the existence of such a direction breaks the manifest
  covariance of the horizons and it is a hindrance to
   construct solutions.

To make further progress, we shall recast the KSEs in a different form using a Dirac operator. In this formulation the $\bbe^\sharp$ direction is not manifest. We shall also present some examples.

\newsection{Horizon Dirac Equation and a Lichnerowicz Theorem}

\subsection{A horizon Dirac Equation}

Given the gravitino KSE in a supergravity theory which is  a parallel transport equation for the supercovariant connection, ${\cal D}_A \e=0$, one can construct a ``supergravity Dirac equation'' as $\Gamma^A {\cal D}_A \e=0$. This can be adapted to the near horizon geometries. In particular, for the ``horizon gravitino KSE'' on ${\cal S}$ ({\ref{sp1}}), which for convenience
can be written as
\bea
\label{redkse1}
\hn_i \phi + \Psi_i \phi =0~,
\eea
where
\bea
\Psi_i = -{1 \over 4} h_i -{1 \over 288} \Gamma_i{}^{\ell_1 \ell_2 \ell_3 \ell_4}
X_{\ell_1 \ell_2 \ell_3 \ell_4} +{1 \over 36} X_{i \ell_1 \ell_2 \ell_3} \Gamma^{\ell_1 \ell_2 \ell_3}
+{1 \over 24} \Gamma_i{}^{\ell_1 \ell_2} Y_{\ell_1 \ell_2} -{1 \over 6} Y_{ij} \Gamma^j~,
\eea
one can define the associated ``horizon Dirac equation'' on ${\cal S}$ as
\bea
\label{dirac1}
\Gamma^i \hn_i \phi + \Psi \phi =0 \ ,
\eea
where
\bea
\Psi = \Gamma^i \Psi_i = -{1 \over 4} h_\ell \Gamma^\ell +{1 \over 96} X_{\ell_1 \ell_2 \ell_3 \ell_4}
\Gamma^{\ell_1 \ell_2 \ell_3 \ell_4} +{1 \over 8} Y_{\ell_1 \ell_2} \Gamma^{\ell_1 \ell_2}~.
\eea
This Dirac equation,  in addition to the Levi-Civita connection, also depends on the fluxes of the supergravity theory restricted on the horizon section ${\cal S}$.

\subsection { A  Lichnerowicz theorem}
The horizon  Dirac equation (\ref{dirac1})  can be used to give a new characterization of the Killing spinors. Clearly, the gravitino KSE is more restrictive and  any solution of the gravitino KSE is also a solution
of a Dirac equation. In what follows, we shall explore the converse. In particular, we shall show that under certain conditions
the zero modes of the horizon Dirac equation are parallel with respect to the horizon supercovariant derivative (\ref{redkse1}).

Before proceeding with the analysis of the supergravity case, it is useful to recall the Lichnerowicz theorem. On any spin compact manifold $N$, one can show the equality
\bea
\int_N \langle \Gamma^i \nabla_i \e, \Gamma^j \nabla_j \e\rangle=  \int_N \langle  \nabla_i \e,  \nabla^i \e\rangle+\int_N {R\over 4} \langle \e, \e\rangle~,
\eea
where $\nabla$ is the Levi-Civita connection, $\langle \cdot, \cdot\rangle$ is the Dirac inner product and $R$ is the Ricci scalar. Clearly if $R>0$, the Dirac operator has no zero modes. Moreover, if
$R=0$, then the zero modes of the Dirac operator are parallel with respect to the Levi-Civita connection.

This theorem can be generalized for M-horizons with the standard Dirac operator replaced with the horizon Dirac operator in (\ref{dirac1}) and the Levi-Civita covariant derivative
replaced with that of the horizon supercovariant derivative  (\ref{redkse1}). For this, let $\phi$ be a Majorana $Spin(9)$ spinor and consider
\bea
\label{lich1}
{\cal{I}} = \int_{\cal{S}} \langle \hn_i \phi + \Psi_i \phi , \hn^i \phi + \Psi^i \phi \rangle
- \int_{\cal{S}} \langle \Gamma^i \hn_i \phi + \Psi \phi, \Gamma^j \hn_j \phi + \Psi \phi \rangle~,
\eea
where $ \langle\cdot, \cdot \rangle$ is the Dirac inner product of $Spin(9)$ which in turn is identified
with the standard Hermitian inner product on $\Lambda^*(\bC^4)$. Observe that $ \langle\cdot, \cdot \rangle$ is positive definite and under this inner product the $Spin(9)$ gamma matrices are Hermitian.


We assume that ${\cal{S}}$ is compact and without boundary, and that  $\phi$ is globally well-defined
and smooth on ${\cal{S}}$.
Then, on integrating by parts, one can rewrite
\begin{eqnarray}
\label{lichaux}
{\cal{I}} &=& \int_{\cal{S}} \langle \phi, (\Psi^{i \dagger} - \Psi^i-(\Psi^\dagger-\Psi)\Gamma^i) {\tilde{\nabla}}_i \phi
+ (\Psi_i^\dagger \Psi^i - \Psi^\dagger \Psi) \phi + \Gamma^{ij} {\tilde{\nabla}}_i {\tilde{\nabla}}_j \phi
\nonumber \\
&+&(\Gamma^i {\tilde{\nabla}}_i \Psi - ({\tilde{\nabla}}^i \Psi_i)) \phi + (\Gamma^i \Psi- \Psi \Gamma^i) {\tilde{\nabla}}_i \phi \rangle~.
\end{eqnarray}
Next, evaluating the RHS of the above equation using the Bianchi identity of $X$ (\ref{clos}), the field equation of the 4-form field strength ({\ref{geq1}}) and ({\ref{geq2}}),
the Einstein equations along ${\cal{S}}$ ({\ref{ein1}})  and taking $\langle \phi, \phi \rangle = \mathrm {const}${\footnote{
This imposes a condition on the spinors which is not usual in the context of a Lichnerowicz theorem. However, it can be shown that from the KSEs side it always holds irrespective of the
identification of $V=W$. Alternatively, the $\langle \phi, \phi \rangle = \mathrm {const}$ condition from the Dirac  side can be removed  provided one in addition imposes $\hn^ih_i=0$ on the geometry.},
one finds that
\bea
\label{lich2}
{\cal{I}} = {\rm Re \ } \bigg( \int_{\cal{S}} \langle \phi , \big( {1 \over 2} h_\ell \Gamma^\ell
-{1 \over 144} X_{\ell_1 \ell_2 \ell_3 \ell_4} \Gamma^{\ell_1 \ell_2 \ell_3 \ell_4}
+{1 \over 6} Y_{\ell_1 \ell_2} \Gamma^{\ell_1 \ell_2} \big)
(\Gamma^i \hn_i \phi + \Psi \phi) \rangle \bigg)~.
\eea
Further details of this computation are given in Appendix C.
On comparing ({\ref{lich2}}) with ({\ref{lich1}}), one immediately finds
that if $\phi$ is a solution of the horizon Dirac equation, then $\phi$ is a solution of the horizon gravitino KSE  (\ref{redkse1}).

Clearly, the above theorem gives a different characterization of supersymmetric horizons. In particular,
if one assumes the Bianchi identity for $X$ ({\ref{clos}}), the 4-form field equations ({\ref{geq1}}) and ({\ref{geq2}}), and
the Einstein equations along ${\cal{S}}$ ({\ref{ein1}}) are satisfied, then the remaining conditions imposed by
supersymmetry and the identification $V=W$ in section \ref{VeW} are ({\ref{blmatch3}}) and the Dirac equation ({\ref{dirac1}}) with
$\phi=1+e_{1234}$. Thus under these conditions the horizon KSE (\ref{redkse1}) can be replaced by the horizon Dirac equation.  We remark that the only condition involving $\Delta$ is ({\ref{blmatch3}}), which therefore defines
$\Delta$ in terms of the other near-horizon data.

\subsection{Solution of horizon Dirac Equation}

As we have demonstrated in the previous section instead of solving the KSEs, it suffices to solve the
the condition ({\ref{blmatch3}}) and the Dirac equation ({\ref{dirac1}}) for $\phi=1+e_{1234}$.
The spinor $\phi=1+e_{1234}$ defines a $Spin(7)$ structure on ${\cal S}$ with  $Spin(7)$ fundamental forms
\bea
\bbe^\sharp~,~~~~\psi = {1 \over 2} (\chi + {\bar{\chi}}) -{1 \over 2} \omega \wedge \omega~,
\eea
where $\chi$ is a $(4,0)$ form, and $\omega$ is an (almost) Hermitian form, see (\ref{hermform}). $Spin(7)$ acts with the ${\bf 1}\oplus {\bf 8}$ representation
on the typical fibre of $T{\cal S}$, and $\bbe^\sharp$ is along the trivial representation while $\psi$ is the fundamental $Spin(7)$ form in the eight directions transverse to $\bbe^\sharp$.

The solution of the linear system obtained from ({\ref{dirac1}}) with $\phi=1+e_{1234}$ can be written as
\bea
h = i_{e_\sharp} Z +{1 \over 2} \theta_\psi - \star_9 (X \wedge \psi)
-{3 \over 2} (\star_9 d \star_9 \bbe^\sharp) \bbe^\sharp~,
\eea
and
\bea
Z^{\bf{7}} = i_\psi \bigg( {1 \over 48} \star_9 d \psi -{1 \over 12} X \bigg)^{\bf{7}}~,
\eea
where
\bea
\label{lfrm}
Z= Y - d \bbe^\sharp, \qquad
\theta_\psi = \star_9 \big( \psi \wedge \star_9 d \psi \big)~,
\eea
ie $\theta_\psi$ is the Lee form of $\psi$.
In addition we have used that if $B \in \Lambda^4 ({\cal{S}})$ is a 4-form, then $i_\psi B \in \Lambda^2 ({\cal{S}})$
with
\bea
(i_\psi B)_{ij} = B_{\ell_1 \ell_2 \ell_3 [i} \psi^{\ell_1 \ell_2 \ell_3}{}_{j]}~,
\eea
and that the superscript ${\bf{7}}$ denotes a projection on the 7-dimensional irreducible representation of $Spin(7)$  in the directions orthogonal to $\bbe^\sharp$.

It is clear that the solution of the horizon Dirac equation can be cast in a more compact form than the
solution to the KSEs. However now the field equations have a more significant role in the construction of solutions.
Nevertheless even though the field equations are second order, when they are restricted  on the near horizon geometries they
take a rather manageable form,  which we shall use to construct solutions.

\section{Solutions}

\subsection{Maximum principle}

In many supergravity theories, the field equations for near horizon geometries are solved  using the maximum principle
 on either $\Delta$  or
$h^2$.  As a consequence, the horizons have either $\Delta$ or $h^2$ constant. As we have already mentioned, it is known
 that there are M-horizons for which $h^2$ is not constant. Nevertheless, it is instructive to seek conditions such that the maximum principle applies  to  either of the two scalars.
To make the application of the maximum principle tangible on $\Delta$ or $h^2$, one has to construct
a second order operator acting on these two functions on ${\cal S}$. A direct application of the
field equations of the theory reveals that
\bea
\tilde\nabla^2\Delta=3 h^i \partial_i\Delta+2\Delta^2-\Delta h^2-{1\over 3}\Delta Y^2- {1\over 72}\Delta X^2-{1\over2} (dh)^2+{1\over6} (d_hY)^2~,
\la{mp1}
\eea
and
\bea
\tilde\nabla^2h^2&=&2 \tilde\nabla_{(i} h_{j)} \tilde\nabla^{(i} h^{j)} +{1\over2}( dh)^2+4 \Delta h^2+h^i\tilde\nabla_i h^2+ (h^2)^2
\cr
&-&{1\over3} h^i X_{ij_1j_2j_3} d_h Y^{j_1j_2j_3}+ h^i d_hY_{ij_1j_2} Y^{j_1j_2}-h^iY_{ik} h^j Y_{j}{}^k+{1\over6} h^i X_{ik_1k_2k_3} h^j X_j{}^{k_1k_2k_3}
\cr
&+&h^2 ({1\over6} Y^2-{1\over72} X^2)-{2\over3} h^i \tilde\nabla_i Y^2-{1\over36} h^i \tilde\nabla_i X^2~.
\la{mp2}
\eea
Observe that the signs of the various terms are indefinite and as a result the maximum principle
does not apply. In what follows, we shall make some simplifying assumptions, and together with
the solutions of the KSEs we shall bring the above equations into a form
which the maximum principle can be applied to.

\subsection{Magnetic solutions}

\subsubsection{The geometry of ${\cal S}$}

For magnetic solutions,  $Y=0$.  In addition, let us take  $\hn^i h_i=0$. The field equation (\ref{einpm}) implies that
\bea
2 \Delta + h^2
-{1 \over 72} X^2=0~.
\eea
Substituting this into (\ref{mp1}), one finds
\bea
{1 \over 2} \hn^2 \Delta -{3 \over 2} h^i \hn_i \Delta = - \Delta h^2 - dh_{ij} dh^{ij} \ .
\eea
Integrating both sides of this equation over ${\cal{S}}$ the contribution from the LHS vanishes. As $\Delta \geq 0$, see ({\ref{blmatch3}}),
this implies that both terms on the RHS must also vanish, ie
\bea
dh=0
\eea
and
\bea
\Delta h =0 \ .
\eea
For the latter condition, there are two possibilities. If $h=0$ then ({\ref{einpi}}) implies
that $\Delta$ is constant. However, we shall concentrate on the second case\footnote{Observe that
if $\Delta=0$, then (\ref{mp1}) is automatically satisfied without other assumptions as a consequence
of  (\ref{q1}).}, for which
$\Delta=0$.

Next turning to (\ref{mp2}), one finds that
\bea
\hn^2 h^2  +2 h^i \hn_i h^2 = 2 \hn^{(i} h^{j)} \hn_{(i} h_{j)} + {1 \over 12} h^i X_{i \ell_1 \ell_2 \ell_3}
h^j X_j{}^{\ell_1 \ell_2 \ell_3}~.
\eea
An application of the maximum principle implies that $h^2$ is constant and both terms on the RHS
must  vanish. Hence, we find that $h$ is covariantly constant on ${\cal{S}}$
\bea
\hn_i h_j=0~.
\eea
We take $h^2 \neq 0$, as if $h=0$ then ({\ref{einpm}}) implies $X=0$ and so $F=0$.
In addition, we find the condition
\bea
\label{inp}
i_h X =0~,
\eea
which together with the Bianchi identity ({\ref{clos}}) implies that
\bea
\cL_h X=0~.
\la{inp2}
\eea

The necessary and sufficient conditions imposed by supersymmetry are obtained from considering the Dirac equation
as described in the previous section. In this case,  ({\ref{blmatch3}}) with $\Delta=0$ implies
the algebraic condition
\bea
\label{cq1}
\bigg( h_\ell \Gamma^\ell +{1 \over 72} X_{\ell_1 \ell_2 \ell_3 \ell_4} \Gamma^{\ell_1 \ell_2 \ell_3 \ell_4} \bigg) \phi =0~,
\eea
and the Dirac equation then simplifies to
\bea
\label{cq2}
\Gamma^i \hn_i \phi - h_i \Gamma^i \phi =0~.
\eea
Thus $\phi$ can be thought of as a spinor coupled to a magnetic gauge potential $h$.

We remark that these conditions are also sufficient to imply the Einstein equations ({\ref{ein1}}), provided that the
condition imposed on the Ricci scalar by ({\ref{ein1}}) is given. To see this, note that the reasoning
used to prove the Lichnerowicz identity set out in Appendix C only makes use of the expression for the Ricci scalar,
and hence it follows that the above conditions are sufficient to imply equation ({\ref{redx3}}). Then, on substituting
the above conditions into ({\ref{redx3}}), one obtains the Einstein equations ({\ref{ein1}}).

For the near horizon Dirac equation to imply the horizon KSE, one has also to  impose the condition $\langle \phi, \phi \rangle =\mathrm{const}$. However in this case, it is not necessary as $\hn^i h_i=0$ and the partial integration argument used to prove the Lichnerowicz identity in appendix C goes through without this condition.  Nevertheless,
one can show that $\langle \phi, \phi \rangle =\mathrm{const}$. To see this, observe that the near horizon Dirac equation (\ref{cq2}) implies the near horizon KSE (\ref{sp1}), and after simplifying the latter using ({\ref{cq1}}), we get
\bea
\label{sp1bb}
\hn_i \phi + \big( {1 \over 4} \Gamma_i{}^\ell h_\ell +{1 \over 24} X_{i \ell_1 \ell_2 \ell_3} \Gamma^{\ell_1 \ell_2 \ell_3}
\big) \phi =0~.
\eea
This in turn gives
\bea
\hn_i \langle \phi, \phi \rangle =0~,
\eea
and hence,  $\langle \phi, \phi \rangle =\mathrm{const}$.
Furthermore on contracting ({\ref{sp1bb}}) with $h$, one obtains
\bea
\label{extracc1}
h^i \hn_i \phi =0 \ .
\eea

Let us summarize the results so far. It is clear that ${\cal S}$ is locally a product ${\cal{S}}= S^1 \times M^8$, where $S^1$ is along the $h$ direction and  $M^8$ is a compact 8-dimensional manifold. Moreover,
({\ref{inp}}) and ({\ref{inp2}}) imply that $X$ is a 4-form on $M^8$. In addition, since $\phi$ does not
depend of the coordinate of $S^1$, all the KSEs and the field equations reduce to equations on $M^8$.

Furthermore  adapting a local co-ordinate $x$ such that $h=dx$, one finds that the
11-dimensional spacetime metric is
\bea
ds^2 = ds_3^2 + ds^2 (M^8)~,
\eea
where
\bea
ds_3^2 = 2 du(dr+r dx) + {dx^2 \over m^2}~,
\eea
is the metric on $AdS_3$, with scalar curvature $R_{AdS_3}=-{3 m^2 \over 2}$, and $m$ is the inverse
radius of $S^1$.
The spacetime is a direct product $AdS_3 \times M^8$.

\subsubsection{Field and KSEs on $M_8$}

We have shown that locally ${\cal S}=S^1\times M^8$ and that the metric can be locally written as
\bea
ds^2({\cal S})= (\bbe^9)^2+ ds^2(M^8)~,~~~\bbe^9={dx\over m}~.
\eea
Thus $h=m\bbe^9$ and $\mathrm{dvol}({\cal S})=\bbe^9\wedge \mathrm{dvol}(M^8)$

Since the flux $X$ is a 4-form on $M^8$ it can be decomposed into self-dual and anti-self dual parts as  $X=X^+ + X^-$. We also decompose the spinor $\phi$ as $\phi=\phi^++\phi^-$ using the projection{\footnote{Note
that here $\pm$ appear as superscripts on spinors, in contrast to subscript $\pm$ elsewhere, which
denote chirality with respect to $\Gamma_{+-}$. Throughout this section, all spinors $\phi$
satisfy $\Gamma_{+-} \phi = \phi$.}} $\Gamma^9 \phi^\pm = \pm \phi^\pm$. This can be identified with the decomposition of Majorana spinors of $Spin(8)$ into chiral and anti-chiral Majorana-Weyl spinors.

The conditions that we have found on ${\cal S}$ imposed by the field and KSEs can now be re-expressed
 as conditions on $M^8$. In particular, the field equations and Bianchi identities reduce to
\bea
\label{ux1}
dX^\pm &=&0 , \qquad (X^\pm)_{A_1 A_2 A_3 A_4} (X^\pm)^{A_1 A_2 A_3 A_4} = 36m^2,
\cr
\label{einr8}
{\tilde{R}}^{(8)}_{AB}& =& -{m^2 \over 2} g_{AB} +{1 \over 12} X_{A N_1 N_2 N_3} X_B{}^{N_1 N_2 N_3}~,~~~\qquad {\tilde{R}}^{(8)} = 2m^2~,
\eea
where ${\tilde{R}}^{(8)}_{AB}$ and ${\tilde{R}}^{(8)}$ is the Ricci tensor\footnote{
Throughout this section, $A,B=1, \dots 8$ are indices of $M^8$ and should not be confused with
the spacetime indices  in section 2.}   and scalar of $M^8$, respectively.
The necessary and sufficient conditions for supersymmetry are
\bea
\label{ux2}
{1 \over 72} (X^\pm)_{A_1 A_2 A_3 A_4} \Gamma^{A_1 A_2 A_3 A_4} \phi^\pm \pm m \phi^\pm =0~,
\eea
and
\bea
\label{neutr}
\Gamma^A {\tilde{\nabla}}_A \phi^\pm  \pm m \phi^\mp =0 \ .
\eea
Moreover the gravitino equation ({\ref{sp1}}) can be written on $M_8$
\bea
\label{gravr8}
{\tilde{\nabla}}_A \phi^\pm \mp {m \over 4} \Gamma_A \phi^\mp +{1 \over 24}
X_{A N_1 N_2 N_3} \Gamma^{N_1 N_2 N_3} \phi^\mp =0 \ .
\eea
We have shown that not all conditions in ({\ref{einr8}}), ({\ref{ux2}}) and ({\ref{neutr}}) and ({\ref{gravr8}}) are independent but it is convenient for the analysis that follows to state them explicitly.

\subsubsection{Topology of $M^8$}

As we have mentioned the existence of a parallel spinor on ${\cal S}$ would imply that ${\cal S}$
admits a nowhere vanishing spinor. This does not impose
a priori  a topological restriction on ${\cal S}$ as all 9-dimensional manifold admit such spinors.

Next let us turn to investigate this question for $M^8$. The relevant parallel transport
equation is (\ref{gravr8}). Again, the existence of a parallel, and so nowhere vanishing,
spinor $\phi=\phi^++\phi^-$ does not impose a priori a topological restriction on $M^8$
as the rank of the spinor bundle is 16 much larger than the dimension of $M^8$. However,
this conclusion holds provided that both components $\phi^\pm$ of $\phi$  do not vanish.
In fact, they are allowed to vanish at subsets of $M^8$ but they cannot vanish everywhere.
To see this, suppose that $\phi^-$ vanishes everywhere. In such a case
\bea
{\tilde{\nabla}}_A \phi^+=0
\eea
and so $M^8$ is a holonomy $Spin(7)$ manifold, ie $\mathrm{hol}(\tilde{\nabla})\subseteq Spin(7)$.
In particular, it will be Ricci flat and so the field equations (\ref{einr8}) cannot be satisfied with $m\not=0$.

The existence of a non-vanishing Majorana-Weyl spinor on a 8-dimensional manifold
imposes a topological restriction, see \cite{isham}. In particular, it is required that
\bea
\pm e-{1\over 2} p_2+{1\over8}p_1^2=0~,
\la{obs}
\eea
where $e$ is the Euler class and $p_2$ and $p_1$ are the Pontryagin classes, and the sign depends
on the chirality of the non-vanishing spinor. We shall explicitly  show  that this condition
is not satisfied for an  example we shall present below.

\subsubsection{Geometry of $M^8$}

In order to evaluate the conditions imposed on the geometry of $M_8$,
note that ({\ref{gravr8}}) implies that
\bea
\label{oneder}
{\tilde{\nabla}}_A f^2 = {m \over 2} \langle \phi^+ , \Gamma_A \phi^- \rangle +{1 \over 12}
X_{A N_1 N_2 N_3} \langle \phi^- , \Gamma^{N_1 N_2 N_3} \phi^+ \rangle~,
\eea
and on taking the divergence, one obtains
\bea
\label{lapv1}
{\tilde{\nabla}}^2 f^2 = m^2 \big( 1-2  f^2 \big)~,
\eea
where we have set $f^2= \langle \phi^+, \phi^+ \rangle$ and
 have adopted the convention $ \langle \phi^+, \phi^+ \rangle+ \langle \phi^-, \phi^- \rangle=1$.
Again, from this equation, we note that $f=0$ and $f=1$ are not solutions, ie  ,
there are no solutions for which either $\phi^+$ or $\phi^-$ vanish identically as has been previously observed.

To proceed further, define the (real) 1-form spinor bilinear $\xi$ by
\bea
\label{xidef}
\xi_A = \langle \phi^- , \Gamma_A \phi^+ \rangle~,
\eea
and it will be convenient to note the identities
\bea
\label{fierz1}
\xi^A \Gamma_A \phi^\pm = \langle \phi^\pm , \phi^\pm \rangle \phi^\mp~,
\eea
and
\bea
\label{fierz2}
\xi^2 = f^2(1-f^2)~.
\eea

Next, we use  ({\ref{gravr8}}) to take the covariant derivative of $\xi$, to obtain
\bea
\label{cder1}
{\tilde{\nabla}}_A \xi_B = -{m \over 4} (2f^2-1) \delta_{AB} +{1 \over 24} X_{A N_1 N_2 N_3}
\big( \langle \phi^+, \Gamma^{N_1 N_2 N_3}{}_B \phi^+ \rangle
+  \langle \phi^-, \Gamma^{N_1 N_2 N_3}{}_B \phi^- \rangle \big). \
\eea
Furthermore, note that ({\ref{ux2}}}) implies that
\bea
X_{A N_1 N_2 N_3} \Gamma^{N_1 N_2 N_3} \phi^\pm =-{1 \over 4} \Gamma_A{}^{N_1 N_2 N_3 N_4}
\phi^\pm \mp 18m \Gamma_A \phi^\pm~,
\eea
which in turn implies that
\bea
X_{N_1 N_2 N_3 [A} \langle \phi^\pm , \Gamma^{N_1 N_2 N_3}{}_{B]} \phi^\pm \rangle =0~.
\eea
On substituting this condition back into ({\ref{cder1}}) one finds that
\bea
\label{closv1}
d \xi =0 \ .
\eea
Further useful identities can be obtained by multiplying ({\ref{ux2}}) with
$\langle \phi^\mp, \phi^\mp \rangle$ and using ({\ref{fierz1}}) to obtain
\bea
{1 \over 72} X_{N_1 N_2 N_3 N_4} \Gamma^{N_1 N_2 N_3 N_4} \xi_B \Gamma^B \phi^\mp
\pm m \langle \phi^\mp , \phi^\mp \rangle \phi^\pm =0~.
\eea
On comparing this expression with that obtained by acting on ({\ref{ux2}}) with $\xi^B \Gamma_B$,
one finds
\bea
\label{xcc1}
(i_\xi X)_{N_1 N_2 N_3} \Gamma^{N_1 N_2 N_3} \phi^\pm = \mp 18m \langle \phi^\pm , \phi^\pm \rangle
\phi^\mp~.
\eea
In addition, on replacing $X$ with $\star_8 X$ and making the appropriate sign changes,
using the same reasoning one finds
\bea
\label{xcc2}
(i_\xi \star_8 X)_{N_1 N_2 N_3} \Gamma^{N_1 N_2 N_3} \phi^+ =0 \ .
\eea

Using these identities, on contracting ({\ref{cder1}}) with $\xi^B$, and making use of the closure of $\xi$,
and ({\ref{fierz2}}), the condition
\bea
\label{extd1}
(2f^2-1) \bigg( m \xi + df^2 \bigg) =0 \ .
\eea
On taking the norm of this expression one finds
\bea
\label{normeq1}
(2f^2-1)^2 \big( {\tilde{\nabla}}_A f^2 {\tilde{\nabla}}^A f^2 - m^2 f^2(1-f^2) \big) =0~.
\eea
We remark however that $f^2={1 \over 2}$ is not a solution, to see this,
contract ({\ref{oneder}}) with $\xi^A$ to obtain
\bea
\label{normeq2}
\xi^A {\hat{\nabla}}_A f^2 = -m f^2(1-f^2)~.
\eea

Additional geometric conditions are obtained by defining the (real) 4-form spinor bilinear
\bea
{\hat{\varphi}}_{A_1 A_2 A_3 A_4} = \langle \phi^+, \Gamma_{A_1 A_2 A_3 A_4} \phi^+ \rangle~.
\eea
Using ({\ref{gravr8}}) and ({\ref{ux2}}) one finds that
\bea
\label{notspin7}
{\tilde{\nabla}}^N {\hat{\varphi}}_{N N_1 N_2 N_3} &=&-{7 \over 2} m \langle \phi^- , \Gamma_{N_1 N_2 N_3}
\phi^+ \rangle -{3 \over 2} (i_\xi X)_{N_1 N_2 N_3}
\nonumber \\
&+& {1 \over 4} \langle \phi^- , X_{M_1 M_2 M_3 [N_1} \Gamma_{N_2 N_3]}{}^{M_1 M_2 M_3} \phi^+ \rangle
\nonumber \\
&-& {3 \over 2} \langle \phi^- , X_{M_1 M_2 [N_1 N_2} \Gamma_{N_3]}{}^{M_1 M_2} \phi^+ \rangle~.
\eea
The final two terms in this expression may be simplified further by noting that
\bea
\label{xxsimp1}
\langle \phi^- , X_{M_1 M_2 [N_1 N_2} \Gamma_{N_3]}{}^{M_1 M_2} \phi^+ \rangle
= {2 \over 3} (i_\xi \star_8 X)_{N_1 N_2 N_3}~,
\eea
and
\bea
\label{xxsimp2}
 \langle \phi^- , X_{M_1 M_2 M_3 [N_1} \Gamma_{N_2 N_3]}{}^{M_1 M_2 M_3} \phi^+ \rangle =
 6m \langle \phi^- , \Gamma_{N_1 N_2 N_3} \phi^+ \rangle +2 (i_\xi X)_{N_1 N_2 N_3}~,
 \eea
 where these conditions follow from  ({\ref{ux2}}) on evaluating
 \bea
 \langle \phi^- , X_{M_1 M_2 M_3 M_4} \Gamma^{M_1 M_2 M_3 M_4} \Gamma_{N_1 N_2 N_3} \phi^+ \rangle~,
 \eea
 and also
  \bea
 \langle \phi^- , \Gamma_{N_1 N_2 N_3} X_{M_1 M_2 M_3 M_4} \Gamma^{M_1 M_2 M_3 M_4}  \phi^+ \rangle~,
 \eea
 and taking the sum and difference of the resulting equations.
 On substituting ({\ref{xxsimp1}}) and ({\ref{xxsimp2}}) into ({\ref{notspin7}}), it follows that
\bea
\label{notspin7b}
{\tilde{\nabla}}^N {\hat{\varphi}}_{N N_1 N_2 N_3} &=&-2 m \langle \phi^- , \Gamma_{N_1 N_2 N_3}
\phi^+ \rangle -2 (i_\xi X^+)_{N_1 N_2 N_3}~.
\eea
It then follows from ({\ref{notspin7b}}) that
\bea
\label{spin7a}
\xi \wedge d {\hat{\varphi}} =0~.
\eea
In addition, using the identity
\bea
\langle \phi^-, \Gamma_{N_1 N_2 N_3} \phi^+ \rangle {\hat{\varphi}}^{N_1 N_2 N_3}{}_M = -42 \langle \phi^+, \phi^+ \rangle \xi_A~,
\eea
together with ({\ref{xcc1}}) and ({\ref{xcc2}}) it follows that
\bea
\label{spin7b}
\star_8 \big( {\hat{\varphi}} \wedge \star_8 d {\hat{\varphi}} \big) = 11mf^2\xi~,
\eea
which relates the Lee form of $\hat\varphi$ to $\xi$.

To summarize, the conditions on the geometry of $M_8$ are given by ({\ref{lapv1}}),
({\ref{closv1}}), ({\ref{extd1}}), ({\ref{normeq1}}), ({\ref{normeq2}}), ({\ref{spin7a}})
and ({\ref{spin7b}}). We remark that if we assume that all spinor bilinears are analytic on
$M_8$, then ({\ref{closv1}}) and ({\ref{normeq2}}) are implied by the other conditions on the
geometry of $M_8$. The Killing spinor equations also determine a number of the components of
$X$ in terms of the geometry. These can be found in appendix D together with a
description of the conditions on the geometry.

\subsubsection{An example}

To construct an explicit example, we take
$M^8 = S^2 \times M^6$, where $M^6$ is a compact K\"ahler 6-manifold.
Next suppose that
$\omega_1$ and $\omega_2$ are the K\"ahler forms of $S^2$ and $M^6$, respectively.
Define
\bea
X^\pm  = {m \over 2} \big( \omega_1 \wedge \omega_2 \pm {1 \over 2} \omega_2 \wedge \omega_2 \big)~.
\eea
It is clear that $X^\pm$ are closed, and that the second algebraic condition in ({\ref{ux1}}) is
also satisfied.

Furthermore to solve the conditions on the spinors $\phi^\pm$ given in ({\ref{ux2}}), we write the spinors $\phi^\pm$ as
\bea
\phi^+ = \eta^{(++)}+ \eta^{(--)}, \qquad \phi^- = \eta^{(+-)}+ \eta^{(-+)}~,
\eea
and restrict them as follows
\bea
{1 \over 2} (\omega_1)_{ab} \Gamma^{ab} \eta^{(\pm \sigma)} = \pm i \eta^{(\pm \sigma)}~,
\qquad
{1 \over 2} (\omega_2)_{rs} \Gamma^{rs} \eta^{(\sigma \pm)} = \pm 3 i \eta^{(\sigma \pm)}~,
\la{sproj}
\eea
where we have decomposed the indices $A=(a, r)$ according  to the product $M^8=S^2\times M^6$
and $\sigma=\pm$. Observe the Clifford algebra element associated with $\omega_2$ has
eigenvalues $\pm 3i$, where the sign depends on the 6-dimensional chirality of the spinor.  So
the  restriction on  the spinors $\phi^\pm$ is that they should lie on the eigenspaces with $\pm 3i$ eigenvalues.
To do this, we have  complexified the spinors to solve the eigenvalue problem but at the end
since we are considering products of projections the resulting eigenspaces are real.

Taking   $X^\pm$ as in ({\ref{ux2}}) and
 spinors satisfying the projection (\ref{sproj}), one can show that the KSEs
can be  solved using
\bea
(\omega_2 \wedge \omega_2)_{A_1 A_2 A_3 A_4} \Gamma^{A_1 A_2 A_3 A_4}
\eta = -144 \eta~,~~
(\omega_1 \wedge \omega_2)_{A_1 A_2 A_3 A_4} \Gamma^{A_1 A_2 A_3 A_4} \eta
= \mp 72 \sigma \eta~,
\eea
where $\sigma$ is the sign of the projection associated with $\omega_1$.

Next, consider the Einstein equations ({\ref{einr8}}).  One finds that
with the  choice of $X^\pm$ as in ({\ref{ux2}}),
${\tilde{R}}^{(6)}_{rs}=0$, which in turn  implies that $M^6$ must be a Calabi-Yau manifold.
Furthermore the Ricci scalar of the $S^2$ must be ${\tilde{R}}^{(2)}_{S^2}=2m^2$, and so $S^2$
is the round 2-sphere.
Hence this solution is $AdS_3 \times S^2 \times \mathrm{CY}^6$, which corresponds to
the uplift of the $D=5$ magnetic black ring near horizon geometry to 11 dimensions.

Let us now investigate the obstruction to the existence of a $Spin(7)$ structure (\ref{obs})
 on $M^8=S^2\times \mathrm{CY}^6$ which arises in the example above. If the obstruction does not vanish,
 it will exclude all  $Spin(7)$ structures on $M^8$ and not only the holonomy $Spin(7)$ structure. For all product manifolds $M^2\times M^6$,
the Pontryagin classes $p_1$ and $p_2$ vanish. Therefore the obstruction to the existence
of a $Spin(7)$ structure is the Euler number which for the example above is
$e(M^8)=e(S^2) e(\mathrm{CY}^6)=2 e(\mathrm{CY}^6)$. So if the Euler number of the Calabi-Yau manifold does not vanish,
then $M^8$ cannot admit a $Spin(7)$ structure as expected.

\newsection{Relation to heterotic horizons}

As another example, we shall demonstrate the heterotic horizons \cite{hethor} are included in M-horizons. For this,
we shall use  the well known compactification ansatz
\bea
ds^2_{(11)}= e^{{4\over3}\mathring{\Phi}} dx^2+ e^{-{2\over3}\mathring{\Phi}} d s^2_{(10)}~,~~~F= dx\wedge H~,
\la{asta}
\eea
and identify the $\bbe^\sharp$ direction of Appendix A with $\bbe^\sharp =-e^{{2\over3}\mathring{\Phi}} dx$,
where $\mathring{\Phi}$ is the dilaton.
In addition, the heterotic horizon geometries can be written as
\bea
ds^2_{(10)}= 2 e^+ e^-+ ds^2_{(8)}~,~~~H=d(e^-\wedge e^+)+H_{(8)}~,~~~dH_{(8)}=0~,
\la{hhor}
\eea
where
\bea
e^+=du~,~~~e^-=ds+s h_{(8)}~,
\eea
$(ds^2_{(8)}, H_{(8)})$ is the metric and 3-form torsion of the spatial horizon section ${\cal S}^8$, and  the dilaton $\mathring{\Phi}$ is a function of  ${\cal S}^8$. If the heterotic
horizons preserve at least one supersymmetry \cite{hethor}, then
\bea
dh_{(8)}\in \mathfrak{spin}(7)~,~~~2d\mathring{\Phi}+h_{(8)}=\theta_\psi~,~~~\mathrm{hol}(\hat\nabla_{(8)})\subseteq Spin(7)~,
\la{hsusy}
\eea
where $\theta_\psi$ is the Lee form of the fundamental $Spin(7)$ form $\psi$ and $\hat\nabla_{(8)}$
is the connection with skew-symmetric torsion on the ${\cal S}^8$. The first two conditions
 are consequences of the dilatino KSE while the last is implied by the gravitino KSE. To embed the heterotic
horizons into M-horizons using (\ref{asta}), we  have to recover the above conditions from
the supersymmetry conditions derived for M-horizons in appendix A.

First substituting (\ref{hhor}) into (\ref{asta}), we demonstrate  that the lifting of a heterotic horizon to 11 dimensions is an M-horizon. Beginning
with the metric and after a straightforward calculation, one finds that the lifted heterotic horizon is
an M-horizon with
\bea
r=e^{-{2\over3}\mathring{\Phi}} s~,~~~h=h_{(8)}+e^{-{2\over3}\mathring{\Phi}} de^{{2\over3}\mathring{\Phi}}~,~~~\Delta=0~.
\la{asta2}
\eea
Similarly, the 3-form flux of the heterotic horizons lifts to a 4-form flux of M-horizons
with
\bea
Y=-\bbe^\sharp\wedge h_{(8)}~,~~   X=-e^{-{2\over3}\mathring{\Phi}} \bbe^\sharp\wedge  H_{(8)}~.
\la{asta3}
\eea
It remains to prove that the supersymmetry conditions of the heterotic horizons (\ref{hsusy}) solve
the conditions of M-horizons stated in appendix A for $\Delta=0$.

It is clear from the form of the flux $F$ associated with these M-horizons that the relevant
conditions are (\ref{d1}), (\ref{q3}), (\ref{q4}), (\ref{q10b}) and (\ref{q11b}). All the remaining
conditions are automatically satisfied. First observe that
\bea
dh=dh_{(8)}~,
\eea
and indeed the condition (\ref{q1}) in Appendix A implies that $dh\in \mathfrak{spin}(7)$. Thus the first
condition in (\ref{hsusy}) is compatible with that in Appendix A.

The second condition in (\ref{hsusy}) is equivalent to (\ref{q4}). This can be verified
by a straightforward computation. Observe that the vanishing of $h_\sharp$ is compatible with
 (\ref{q3}) as  the  direction $\bbe^\sharp$
is generated by an isometry.

It remains to demonstrate that the last equation in (\ref{hsusy}) is also equivalent to
the conditions (\ref{q10b}) and (\ref{q11b}) stated in appendix A. For this suffices to show that the gravitino
KSE of M-theory for the backgrounds  given in (\ref{asta}), (\ref{asta2}) and (\ref{asta3}) reduce
to that of heterotic supergravity.  In both cases, the gravitino KSEs have been evaluated
on the same   spinor $\phi=1+e_{1234}$.
To prove this, first act with a gamma matrix on  $\Theta_+\phi=0$, as $\Delta=0$, to find
\bea
\big( h_i+ \Gamma_i{}^j h_j+{1\over72} X_{\ell_1\dots \ell_4} \Gamma_i{}^{\ell_1\dots\ell_4}+{1\over18} X_{i\ell_1\ell_2\ell_3} \Gamma^{\ell_1\ell_2\ell_3}
+{1\over3} Y_{\ell_1\ell_2} \Gamma_i{}^{\ell_1\ell_2}+{2\over3} Y_{ij} \Gamma^j\big)\phi=0~,
\eea
where $i,j,k=1,\dots 8$ are indices of the heterotic horizons section ${\cal S}^8$.
Using this, the parallel transport equation on the horizon section (\ref{redkse1})  can be simplified as
\bea
\hn_i \phi+\big({1\over4} \Gamma_i{}^j h_j+{1\over24} X_{i\ell_1\ell_2\ell_3}\Gamma^{\ell_1\ell_2\ell_3}
+{1\over8} \Gamma_i{}^{\ell_1\ell_2} Y_{\ell_1\ell_2}\big)\phi=0~.
\eea
Evaluating this on the background (\ref{asta}), (\ref{asta2}) and (\ref{asta3}) and along
the $i=(\a, \bar\a)$ directions, we find that
\bea
(\hn_{(8)\,i}-{1\over8} H_{ijk} \Gamma^{jk})\phi=0~,~~~
\eea
where now the frame components of the $H$ flux has been computed in terms of the frame
 of the 10-dimensional metric and $i,j,k$ are frame indices of the heterotic horizon section ${\cal S}^8$. The resulting expression is the gravitino KSE of the heterotic horizons.
 As a result all heterotic horizons can be lifted to M-theory horizons. The 9-dimensional
  sections of M-horizons are warped products, ${\cal S}^9=S^1\times_w {\cal S}^8$,
   \bea
  ds^2({\cal S}^9)= e^{{4\over3}\mathring{\Phi}} dx^2+ e^{-{2\over3}\mathring{\Phi}} d s^2({\cal S}^8)~,
  \eea
   and the warp factors are related to
  the  dilaton.

 We have shown that all heterotic horizons
are special cases of M-horizons. One of the key properties of the heterotic horizons
is that exhibit supersymmetry enhancement. In particular, if $h_{(8)}\not=0$, then the heterotic
horizons preserve at least two supersymmetries. This class of backgrounds also exhibits
supersymmetry enhancement in M-theory. However, it is not apparent that all M-horizons preserve
at least two supersymmetries.

\newsection{Concluding Remarks}

We have identified the geometry of M-horizons that preserve at least one supersymmetry. The horizon
sections are 9-dimensional manifolds with a $Spin(7)$ structure which is appropriately geometrically restricted.
The full solution of the KSEs and the geometric conditions are presented in appendix A. We have also proved, using the field
equations, a Lichnerowicz type of theorem for horizon sections. In particular, we have shown that
the zero modes of an appropriate Dirac equation which couples to the 4-form fluxes are parallel
with respect to the supercovariant connection as restricted to a horizon section. This gives an alternative characterization
of the solution to the KSEs which is advantageous for the investigation of some examples. We have
also shown that the heterotic horizons can be lifted to M-horizons.

Although the KSE for M-horizons can be solved for one Killing spinor and the
geometry of the solutions can be identified, the understanding of these backgrounds is less complete
than those of other supergravity theories, such as  ${\cal N}=1$ theories in 4 and 5 dimensions,  $(1,0)$ 6-dimensional supergravity and heterotic
theories. There are several obstacles to this. The first  obstacle  is  our limited understanding of the supersymmetric
solutions of 11-dimensional supergravity preserving any number of supersymmetries. This is a well
known problem and apart from the solution of the KSEs for one Killing spinor, and the classification
of backgrounds with more than 29 Killing spinors \cite{max, n31, n30}, no other general results are available, see
also \cite{duff}.  The second obstacle is to find a general method to solve the field equations which
are not implied by the KSEs. This is typically done by applying the maximum principle
on a scalar constructed from the data of the problem and utilizing the compactness of the horizon
section. The scalars that are typically used for this are either $\Delta$ or  $h^2$ that appear in the metric (\ref{mhm}). However, we know that for general M-horizons one cannot formulate a maximum principle
for $h^2$ as there are examples for which $h^2$ is not a constant. For $\Delta$ the status of the maximum principle is less
clear. We give the equations, which are derived from the field equations, that restrict $\Delta$ and $h^2$ and could be suitable for
an application of the maximum principle.  But as expected the various terms that enter have an indefinite sign
and so the maximum principle cannot apply unless additional restrictions are imposed on the fields by hand.
Perhaps this problem of applying the maximum principle can be resolved for M-horizons
preserving  sufficiently large number of supersymmetries.

A related problem for M-horizons is supersymmetry enhancement. For many horizons that preserve one supersymmetry one can show, using the compactness
of the horizon sections and an application of the maximum principle,  that the solution
exhibits supersymmetry enhancement. For example heterotic horizons with non-vanishing rotation preserve 2,4,6,8 and 16
supersymmetries. For M-horizons, it is not apparent that such an enhancement takes place. Again,
the role played by the maximum principle in supersymmetry enhancement is expected to be particularly important.

Despite these issues, significant progress has been made in the
identification of M-horizon geometries. It is clear that the geometry
of M-horizons preserving more than one supersymmetry is  a special case of that which we have found
for  horizons preserving one supersymmetry. In addition, the Lichnerowicz type of techniques
we have introduced give a new insight into the topology and geometry of the horizon sections, and lead  to a generalization of the Lichnerowicz theorem. This technique can be applied to similar problems
in other supergravity theories and may lead to a better understanding of
 supergravity backgrounds.

\vskip 0.5cm
\noindent{\bf Acknowledgements} \vskip 0.1cm
\noindent  JG is supported by the STFC grant, ST/1004874/1.
GP is partially supported by the  STFC rolling grant ST/J002798/1.
\vskip 0.5cm

\newpage

 \setcounter{section}{0}

\appendix{Solution of Linear System}

\subsection{Conventions and Spinor bi-linears}

The choice of the spacetime frame and our spinor conventions are as those in \cite{system11}. In particular,
the eleven-dimensional volume form  is chosen as
\bea
d{\rm vol} = \bbe^{0512346789\sharp}~.
\eea
In terms of the complex frame basis $e^\a={1\over\sqrt 2} (\bbe^\a+ i \bbe^{\a+5})$, $\a=1,2,3,4$, adapted to the realization of spinors as multi-forms,  we  have
\bea
d{\rm vol} = \bbe^{+-}\wedge e^{1 \bar{1}} \wedge e^{2 \bar{2}} \wedge
e^{3 \bar{3}} \wedge e^{4 \bar{4}} \wedge \bbe^\sharp~,
\eea
where the spacetime indices decompose as $A=(+,-, i)$ and $i=(\sharp, \a, \bar\a)$.
The volume form can also be expressed in terms of the (almost) Hermitian form $\omega$ and the (4,0)-form $\chi$,
\bea
\omega=-i\delta_{\a\bar\b} e^\a\wedge e^{\bar\b}~,~~~\chi=4 e^{1234}~,~~~
\la{hermform}
\eea
respectively,  as
\bea
d{\rm vol} ={1\over24} \bbe^+ \wedge \bbe^- \wedge \omega \wedge \omega \wedge
\omega \wedge \omega \wedge \bbe^\sharp={1\over16} \bbe^+ \wedge \bbe^- \wedge \chi \wedge {\bar{\chi}} \wedge \bbe^\sharp~.
\eea

A direct computation reveals that the form bilinears of the Killing spinor $\e$ in (\ref{gfixks}) up to an overall numerical normalization are as follows. The 1-form bilinear is
\bea
V= \bbe^--{1\over2} \Delta r^2 \bbe^+ \ .
\eea
The 2-form bilinear is
\bea
\alpha=2 (\bbe^-+{1\over2} r^2 \Delta \bbe^+)\wedge \bbe^\sharp-4 r \Phi \omega
\eea
and the 5-form bilinear is
\bea
\sigma=- \{(\bbe^-+{1\over2} \Delta r^2 \bbe^++2i r\Phi \bbe^\sharp)\wedge \chi+ {\rm c.c.}\}+ (\bbe^--{1\over2} \Delta r^2 \bbe^+)\wedge
\omega\wedge \omega~.
\eea

\subsection{Linear system}
Suppose that we choose the gauge  $\zeta=\Theta_+ \phi_+=i \Phi (1-e_{1234})$, see (\ref{gfix}).
Then the linear system obtained from ({\ref{sp1}}),
({\ref{sp2}}) and ({\ref{gfix}}) is

\be
\label{e1}
-{1 \over 2} \Omega_{\alpha, \beta}{}^\beta +{1 \over 4} h_\alpha -{1 \over 4} X_{\sharp \alpha \beta}{}^\beta +{1 \over 4}
Y_{\sharp \alpha} = 0~,
\ee
\be
\label{e2}
\Omega_{\mu_1, \sharp \mu_2} -{1 \over 2} X_{\mu_1 \mu_2 \lambda}{}^\lambda
-{1 \over 6} X_{\mu_1 {\bar{\lambda}}_1  {\bar{\lambda}}_2
 {\bar{\lambda}}_3} \epsilon^{{\bar{\lambda}}_1  {\bar{\lambda}}_2
 {\bar{\lambda}}_3}{}_{\mu_2} +{1 \over 2} Y_{\mu_1 \mu_2} -{1 \over 2}
 Y_{{\bar{\lambda}}_1  {\bar{\lambda}}_2} \epsilon^{{\bar{\lambda}}_1  {\bar{\lambda}}_2}{}_{\mu_1 \mu_2}=0~,
 \ee
\bea
\label{e3}
{1 \over 2} \Omega_{\beta, \mu_1 \mu_2}
-{1 \over 4} \Omega_{\beta, {\bar{\lambda}}_1  {\bar{\lambda}}_2} \epsilon^{{\bar{\lambda}}_1  {\bar{\lambda}}_2}{}_{\mu_1 \mu_2}
+{1 \over 4} X_{\sharp \beta \mu_1 \mu_2} -{1 \over 8} X_{\sharp \beta {\bar{\lambda}}_1  {\bar{\lambda}}_2}
\epsilon^{{\bar{\lambda}}_1  {\bar{\lambda}}_2}{}_{\mu_1 \mu_2}
+{1 \over 4}(h_{\bar{\lambda}} + Y_{\sharp {\bar{\lambda}}}) \epsilon^{\bar{\lambda}}{}_{\beta \mu_1 \mu_2}=0~,
\nonumber \\
\eea
\be
\label{e4}
-\Omega_{\bar{\beta}, \sharp \alpha} +{1 \over 2} X_{\bar{\beta} \alpha \lambda}{}^\lambda
+{1 \over 2} Y_{\bar{\beta} \alpha}
+ \bigg(-{1 \over 4}h_\sharp +i \Phi +{1 \over 8} X_\sigma{}^\sigma{}_\lambda{}^\lambda \bigg) \delta_{\bar{\beta} \alpha}=0~,
\ee
\be
\label{e5}
\Omega_{\sharp, \lambda}{}^\lambda -{1 \over 2} Y_\lambda{}^\lambda +2i \Phi =0~,
\ee
\be
\label{e6}
\Omega_{\sharp, \sharp \alpha} +{1 \over 2} h_\alpha -{1 \over 2} X_{\sharp \alpha \lambda}{}^\lambda
-{1 \over 6} X_{\sharp {\bar{\lambda}}_1  {\bar{\lambda}}_2
 {\bar{\lambda}}_3} \epsilon^{{\bar{\lambda}}_1  {\bar{\lambda}}_2
 {\bar{\lambda}}_3}{}_\alpha =0~,
 \ee
 \be
 \label{e7}
 {1 \over 2} \Omega_{\sharp, \mu_1 \mu_2} -{1 \over 4} \Omega_{\sharp, {\bar{\lambda}}_1  {\bar{\lambda}}_2}
 \epsilon^{{\bar{\lambda}}_1  {\bar{\lambda}}_2}{}_{\mu_1 \mu_2}
 -{1 \over 4} Y_{\mu_1 \mu_2} +{1 \over 8} Y_{{\bar{\lambda}}_1  {\bar{\lambda}}_2}
 \epsilon^{{\bar{\lambda}}_1  {\bar{\lambda}}_2}{}_{\mu_1 \mu_2}=0~,
 \ee
 \be
 \label{g1}
 -h_\sharp +{2 \over 3} Y_\lambda{}^\lambda +{1 \over 6} X_\sigma{}^\sigma{}_\lambda{}^\lambda
 +{1 \over 18} X_{\mu_1 \mu_2 \mu_3 \mu_4}\epsilon^{\mu_1 \mu_2 \mu_3 \mu_4}
 -4i \Phi =0~,
 \ee
 \be
 \label{g2}
 h_\alpha +{2 \over 3} Y_{\sharp \alpha}-{1 \over 3} X_{\sharp \alpha \lambda}{}^\lambda -{1 \over 9}
 X_{\sharp {\bar{\lambda}}_1  {\bar{\lambda}}_2
 {\bar{\lambda}}_3} \epsilon^{{\bar{\lambda}}_1  {\bar{\lambda}}_2
 {\bar{\lambda}}_3}{}_\alpha =0~,
 \ee
 \be
 \label{g3}
 2 Y_{\mu_1 \mu_2} -Y_{ {\bar{\lambda}}_1  {\bar{\lambda}}_2} \epsilon^{ {\bar{\lambda}}_1  {\bar{\lambda}}_2}{}_{\mu_1 \mu_2}
 -X_\sigma{}^\sigma{}_{\mu_1 \mu_2} -{1 \over 2}X_\sigma{}^\sigma{}_{ {\bar{\lambda}}_1  {\bar{\lambda}}_2}
 \epsilon^{ {\bar{\lambda}}_1  {\bar{\lambda}}_2}{}_{\mu_1 \mu_2}=0~,
 \ee
  \be
  \label{f1}
  -i \hn_\alpha \Phi +{i \over 2} \Phi h_\alpha -{i \over 2} \Phi X_{\sharp \alpha \lambda}{}^\lambda
  -{1 \over 2} dh_{\sharp \alpha} +{1 \over 12} \epsilon_\alpha{}^{{\bar{\lambda}}_1  {\bar{\lambda}}_2
 {\bar{\lambda}}_3} d_hY_{{\bar{\lambda}}_1  {\bar{\lambda}}_2
 {\bar{\lambda}}_3}=0~,
  \ee
 \be
 \label{f2}
 -i \Phi X_{\sharp \alpha \lambda}{}^\lambda -{3 \over 4} dh_{\sharp \alpha}
 -{1 \over 4} d_hY_{\alpha \lambda}{}^\lambda
 + {1 \over 12} \epsilon_\alpha{}^{{\bar{\lambda}}_1  {\bar{\lambda}}_2
 {\bar{\lambda}}_3} d_hY_{{\bar{\lambda}}_1  {\bar{\lambda}}_2
 {\bar{\lambda}}_3}=0~,
\ee
\bea
\label{f3}
-i \Phi Y_{{\bar{\lambda}}_1  {\bar{\lambda}}_2} \epsilon^{{\bar{\lambda}}_1  {\bar{\lambda}}_2}{}_{\mu_1 \mu_2}
- i \Phi X_{\mu_1 \mu_2 \lambda}{}^\lambda
-dh_{\mu_1 \mu_2}
\nn
+{1 \over 4} dh_{{\bar{\lambda}}_1  {\bar{\lambda}}_2} \epsilon^{{\bar{\lambda}}_1  {\bar{\lambda}}_2}{}_{\mu_1
\mu_2} +{1 \over 4}d_hY_{\sharp {\bar{\lambda}}_1  {\bar{\lambda}}_2}
\epsilon^{{\bar{\lambda}}_1  {\bar{\lambda}}_2}{}_{\mu_1 \mu_2}=0~,
\eea
\bea
\label{f4}
-i \Phi \Omega_{\beta, \mu_1 \mu_2}+{i \over 4}\Phi X_{\sharp \beta {\bar{\lambda}}_1  {\bar{\lambda}}_2}
\epsilon^{{\bar{\lambda}}_1  {\bar{\lambda}}_2}{}_{\mu_1 \mu_2} -{1 \over 8}
dh_{\sharp \bar{\lambda}} \epsilon^{\bar{\lambda}}{}_{\beta \mu_1 \mu_2}
\nn
+{1 \over 8} d_hY_\mu{}^\mu{}_{\bar{\lambda}}
\epsilon^{\bar{\lambda}}{}_{\beta \mu_1 \mu_2}
+{1 \over 8}d_hY_{\beta {\bar{\lambda}}_1  {\bar{\lambda}}_2} \epsilon^{{\bar{\lambda}}_1  {\bar{\lambda}}_2}{}_{\mu_1
\mu_2}=0~,
\eea
\bea
\label{f5}
i \Phi X_{\bar{\beta} \alpha \lambda}{}^\lambda
-{1 \over 2} d_hY_{\sharp {\bar{\beta}} \alpha} +{1 \over 2}dh_{\bar{\beta} \alpha}
+ \bigg(-{1 \over 4}dh_\lambda{}^\lambda -{1 \over 4}d_hY_{\sharp \lambda}{}^\lambda \bigg)
\delta_{\bar{\beta} \alpha} =0~,
\eea
\bea
\label{f6}
i \hn_\sharp \Phi -{i \over 2} \Phi h_\sharp +{1 \over 8} dh_\lambda{}^\lambda
-{1 \over 8} d_hY_{\sharp \lambda}{}^\lambda =0~,
\eea
\bea
\label{f7}
-i \Phi \Omega_{\sharp \mu_1 \mu_2} +{i \over 2}\Phi Y_{\mu_1 \mu_2}
+{1 \over 8} dh_{\mu_1 \mu_2} -{1 \over 16} dh_{{\bar{\lambda}}_1  {\bar{\lambda}}_2}
\epsilon^{{\bar{\lambda}}_1  {\bar{\lambda}}_2}{}_{\mu_1 \mu_2}
\nn
-{1 \over 8} d_hY_{\sharp \mu_1 \mu_2}
+{1 \over 16} d_hY_{\sharp {\bar{\lambda}}_1  {\bar{\lambda}}_2}
\epsilon^{{\bar{\lambda}}_1  {\bar{\lambda}}_2}{}_{\mu_1 \mu_2}=0~.
\eea
The above conditions have been expressed in representations of the $S(4)$ subgroup of $Spin(7)$.
Investigating these conditions, one can show   that
the algebraic conditions ({\ref{cc1}}) and ({\ref{cc2}})  hold automatically, as a consequence
of ({\ref{sp1}}), ({\ref{sp2}}) and ({\ref{thetdef}}).

\subsection{Solution for $\Delta\not=0$}

Suppose that $\Delta\not=0$ and so $\Phi\not=0$ at some patch. Under this assumption, the linear system can be solved to express
some of the fluxes in terms of the geometry and determine the conditions on the geometry imposed by supersymmetry.
In particular, the conditions
({\ref{e1}})-({\ref{f7}}) can be solved as
\be
\label{q1}
Y = - d_h \bbe^\sharp -2 \Phi {{\omega}}~,
\ee
where $\omega$ is given in (\ref{hermform}),
and
\be
\label{q2}
\Phi = -{i \over 2} \big( \Omega_{\sharp, \lambda}{}^\lambda+{1 \over 2} \Omega^\lambda{}_{, \sharp \lambda}
-{1 \over 2} \Omega_{\lambda, \sharp}{}^\lambda \big)~,
\ee
\be
\label{q3}
h_\sharp = -{1 \over 2} \big(\Omega^\lambda{}_{,\sharp \lambda} +\Omega_{\lambda, \sharp}{}^\lambda \big)~,
\ee
\be
\label{q4}
h_\alpha = -{2 \over 3} \Omega_{\bar{\lambda}_1 , \bar{\lambda}_2 \bar{\lambda}_3}
\epsilon^{\bar{\lambda}_1 \bar{\lambda}_2 \bar{\lambda}_3}{}_\alpha +{4 \over 3}
\Omega_{\bar{\lambda},}{}^{\bar{\lambda}}{}_\alpha -{2 \over 3} \Omega_{\alpha, \lambda}{}^\lambda
+{1 \over 3} \Omega_{\sharp, \sharp \alpha}~.
\ee
Also, $\Phi$ is given as
\be
\label{q5}
\partial_\alpha \Phi = \Phi \big(-{4 \over 3} \Omega_{\bar{\lambda}_1 , \bar{\lambda}_2 \bar{\lambda}_3}
\epsilon^{\bar{\lambda}_1 \bar{\lambda}_2 \bar{\lambda}_3}{}_\alpha
+{2 \over 3} \Omega_{\bar{\lambda},}{}^{\bar{\lambda}}{}_\alpha  +{2 \over 3}\Omega_{\alpha,\lambda}{}^\lambda
-{1 \over 3} \Omega_{\sharp, \sharp \alpha} \big)+{i \over 2} dh_{\sharp \alpha}~,
\ee
\be
\label{q6}
\partial_\sharp \Phi = -{1 \over 4} \Phi \big( \Omega^\lambda{}_{,\sharp \lambda}+\Omega_{\lambda, \sharp}{}^\lambda
+X_\lambda{}^\lambda{}_\sigma{}^\sigma \big)+{i \over 2} dh_\lambda{}^\lambda~.
\ee
The 4-form $X$ can be expressed
\be
\label{q7}
X_{\mu \bar{\lambda}_1 \bar{\lambda}_2 \bar{\lambda}_3}
= \big(-\Omega_{\mu, \sharp \sigma}+\Omega_{\sharp, \mu \sigma}
-2 \Omega_{[\mu|, \sharp |\sigma]}
+{i \over 2} \Phi^{-1} dh_{\mu \sigma}-{i \over 4} \Phi^{-1} dh_{\bar{\nu}_1 \bar{\nu}_2}
\epsilon^{ \bar{\nu}_1 \bar{\nu}_2}{}_{\mu \sigma} \big) \epsilon^\sigma{}_{ \bar{\lambda}_1 \bar{\lambda}_2 \bar{\lambda}_3}~,
\ee
\be
\label{q8}
X_{\bar{\beta} \alpha \lambda}{}^\lambda + {1 \over 4}  X_\lambda{}^\lambda{}_\sigma{}^\sigma \delta_{\bar{\beta} \alpha}
= \Omega_{\bar{\beta}, \sharp \alpha}+ \Omega_{\alpha, \sharp \bar{\beta}}
-{1 \over 4} \big(\Omega^\lambda{}_{, \sharp \lambda}+\Omega_{\lambda, \sharp}{}^\lambda \big) \delta_{\bar{\beta}
\alpha}~,
\ee
\be
\label{q9}
{2 \over 3} \Omega_{\sharp, \lambda}{}^\lambda +{1 \over 6} \Omega^{\lambda}{}_{, \sharp \lambda}
+{5 \over 6} \Omega_{\lambda, \sharp}{}^\lambda +{1 \over 6}X_\lambda{}^\lambda{}_\sigma{}^\sigma
+{1 \over 18} X_{\lambda_1 \lambda_2 \lambda_3 \lambda_4} \epsilon^{\lambda_1 \lambda_2 \lambda_3 \lambda_4}=0~,
\ee
\be
\label{q10}
X_{\sharp \lambda_1 \lambda_2 \lambda_3} = -2 \Omega_{[\lambda_1, \lambda_2 \lambda_3]}
+{2 \over 3} \big(-\Omega_{\lambda,}{}^\lambda{}_{\bar{\sigma}}+\Omega_{\bar{\sigma}, \lambda}{}^\lambda
-\Omega_{\sharp, \sharp \bar{\sigma}} \big) \epsilon^{\bar{\sigma}}{}_{\lambda_1 \lambda_2 \lambda_3}~,
\ee
\bea
\label{q11}
X_{\sharp \beta \bar{\sigma}_1 \bar{\sigma}_2}
&=& {2 \over 3} \big(\Omega_{\beta, \mu_1 \mu_2}
+ \Omega_{\mu_1, \beta \mu_2} \big) \epsilon^{\mu_1 \mu_2}{}_{\bar{\sigma}_1 \bar{\sigma}_2}
-2 \Omega_{\beta, \bar{\sigma}_1 \bar{\sigma}_2}
\nn
&+& \bigg(-{4 \over 3} \Omega_{\lambda ,}{}^\lambda{}_{[ {\bar{\sigma}_1}}
+{4 \over 3} \Omega_{[\bar{\sigma}_1 , |\lambda|}{}^\lambda
+{2 \over 3} \Omega_{\sharp, \sharp [{\bar{\sigma}}_1} \bigg) \delta_{\bar{\sigma}_2 ] \beta}~.
\eea
The linear system also imposes the following condition on the geometry
\be
\label{q12}
\Omega_{[\mu_1, |\sharp| \mu_2]}-\Omega_{\sharp, \mu_1 \mu_2}
-{1 \over 2} \big(\Omega_{\bar{\sigma}_1, \sharp \bar{\sigma}_2}
-\Omega_{\sharp, \bar{\sigma}_1 \bar{\sigma}_2} \big) \epsilon^{\bar{\sigma}_1 \bar{\sigma}_2}{}_{\mu_1 \mu_2}=0~.
\ee
This is the only constraint on the geometry of ${\cal S}$ directly implied from the solution of the KSEs in the ({\ref{gfix}})
gauge. One can derive  additional conditions  after substituting the expressions of $\Phi$ and $h$ in terms of the geometry into
(\ref{q5}) and (\ref{q6}). Such conditions are quadratic in the spin connection and
are reminiscent of integrability conditions.

\subsection{Solution for $\Delta=0$}

Let us now turn to the  $\Delta=\Phi=0$ case. The solution to the linear system is
\be
\label{q1b}
Y = - d_h \bbe^\sharp~,
\ee
and   $h$ is given as in (\ref{q3}) and (\ref{q4}). Moreover,
\bea
dh_{\sharp \alpha}=dh_\lambda{}^\lambda= 0~,~~~dh_{\a_1\a_2}-{1\over2}\e_{\a_1\a_2}{}^{\bar\b_1\bar\b_2} dh_{\bar\b_1\bar\b_2}=0~.
\la{d1}
\eea

The 4-form $X$ can be expressed
\bea
\label{q7b}
&&X_{(\mu_1 |\bar\lambda_1 \bar\lambda_2 \bar\lambda_3}|}\e^{\bar\lambda_1 \bar\lambda_2 \bar\lambda_3}{}_{\mu_2)}=6 \Omega_{(\mu_1,|\sharp| \mu_2)~,
\cr
&&X_{\a_1\a_2\l}{}^\l+{1\over2} X_{\bar\mu_1\bar\mu_2\l}{}^\l \e^{\bar\mu_1\bar\mu_2}{}_{\a_1\a_2}=4 \Omega_{[\a_1,|\sharp| \a_2]}-2  \Omega_{[\bar\mu_1,|\sharp| \bar\mu_2]}
\e^{\bar\mu_1\bar\mu_2}{}_{\a_1\a_2}~,
\eea

\be
\label{q8b}
X_{\bar{\beta} \alpha \lambda}{}^\lambda + {1 \over 4}  X_\lambda{}^\lambda{}_\sigma{}^\sigma \delta_{\bar{\beta} \alpha}
= \Omega_{\bar{\beta}, \sharp \alpha}+ \Omega_{\alpha, \sharp \bar{\beta}}
-{1 \over 4} \big(\Omega^\lambda{}_{, \sharp \lambda}+\Omega_{\lambda, \sharp}{}^\lambda \big) \delta_{\bar{\beta}
\alpha}~,
\ee
\be
\label{q9b}
 -{1 \over 6} \Omega^{\lambda}{}_{, \sharp \lambda}
+{7 \over 6} \Omega_{\lambda, \sharp}{}^\lambda +{1 \over 6}X_\lambda{}^\lambda{}_\sigma{}^\sigma
+{1 \over 18} X_{\lambda_1 \lambda_2 \lambda_3 \lambda_4} \epsilon^{\lambda_1 \lambda_2 \lambda_3 \lambda_4}=0~,
\ee
\be
\label{q10b}
X_{\sharp \lambda_1 \lambda_2 \lambda_3} = -2 \Omega_{[\lambda_1, \lambda_2 \lambda_3]}
+{2 \over 3} \big(-\Omega_{\lambda,}{}^\lambda{}_{\bar{\sigma}}+\Omega_{\bar{\sigma}, \lambda}{}^\lambda
-\Omega_{\sharp, \sharp \bar{\sigma}} \big) \epsilon^{\bar{\sigma}}{}_{\lambda_1 \lambda_2 \lambda_3}~,
\ee
\bea
\label{q11b}
X_{\sharp \beta \bar{\sigma}_1 \bar{\sigma}_2}
&=& {2 \over 3} \big(\Omega_{\beta, \mu_1 \mu_2}
+ \Omega_{\mu_1, \beta \mu_2} \big) \epsilon^{\mu_1 \mu_2}{}_{\bar{\sigma}_1 \bar{\sigma}_2}
-2 \Omega_{\beta, \bar{\sigma}_1 \bar{\sigma}_2}
\nn
&+& \bigg(-{4 \over 3} \Omega_{\lambda ,}{}^\lambda{}_{[ {\bar{\sigma}_1}}
+{4 \over 3} \Omega_{[\bar{\sigma}_1 , |\lambda|}{}^\lambda
+{2 \over 3} \Omega_{\sharp, \sharp [{\bar{\sigma}}_1} \bigg) \delta_{\bar{\sigma}_2 ] \beta}~.
\eea
The linear system also imposes the following conditions on the geometry
\be
\label{q12b}
\Omega_{[\mu_1, |\sharp| \mu_2]}-\Omega_{\sharp, \mu_1 \mu_2}
-{1 \over 2} \big(\Omega_{\bar{\sigma}_1, \sharp \bar{\sigma}_2}
-\Omega_{\sharp, \bar{\sigma}_1 \bar{\sigma}_2} \big) \epsilon^{\bar{\sigma}_1 \bar{\sigma}_2}{}_{\mu_1 \mu_2}&=&0~,
\cr
 \Omega_{\sharp, \lambda}{}^\lambda+{1 \over 2} \Omega^\lambda{}_{, \sharp \lambda}
-{1 \over 2} \Omega_{\lambda, \sharp}{}^\lambda &=&0~.
\ee
This concludes the analysis of the linear system.

\appendix{Analysis of Integrability Conditions}
In this Appendix we present the proof that ({\ref{sp2}}) is implied
by ({\ref{sp1}}) and ({\ref{thetdef}}), together with  ({\ref{clos}}), the 4-form field equations
({\ref{geq1}}) and ({\ref{geq2}}), and the components of the Einstein equations on ${\cal{S}}$.

First, use ({\ref{thetdef}}) to eliminate the $\zeta$ from ({\ref{sp2}}), and then
use ({\ref{sp1}}) to eliminate the supercovariant derivative of $1+e_{1234}$. Then
({\ref{sp2}}) is equivalent to
\bea
\label{redx1}
&&\bigg( {1 \over 4} \hn_j h_i \Gamma^j +{1 \over 72} \Gamma_i{}^{\ell_1 \ell_2 \ell_3}
dY_{\ell_1 \ell_2 \ell_3} -{1 \over 12} (dY)_{i \ell_1 \ell_2} \Gamma^{\ell_1 \ell_2}
+{1 \over 12} \hn_i Y_{\ell_1 \ell_2} \Gamma^{\ell_1 \ell_2}
\nn
&&+{1 \over 288} \hn_i X_{\ell_1 \ell_2 \ell_3 \ell_4} \Gamma^{\ell_1 \ell_2 \ell_3 \ell_4}
-{1 \over 48} \Gamma_i{}^{\ell_1 \ell_2 \ell_3} h_{\ell_1} Y_{\ell_2 \ell_3}
+{1 \over 8} h_{[i} Y_{\ell_1 \ell_2]} \Gamma^{\ell_1 \ell_2}
\nn
&&-{1 \over 144} (i_h X)_{\ell_1 \ell_2 \ell_3} \Gamma_i{}^{\ell_1 \ell_2 \ell_3}
+{1 \over 24} (i_h X)_{i \ell_1 \ell_2} \Gamma^{\ell_1 \ell_2} -{1 \over 8} h_i h_j \Gamma^j
\nn
&&+{1 \over 41472} \Gamma_i{}^{\ell_1 \ell_2 \ell_3 \ell_4 \ell_5 \ell_6 \ell_7 \ell_8}
X_{\ell_1 \ell_2 \ell_3 \ell_4} X_{\ell_5 \ell_6 \ell_7 \ell_8}
-{1 \over 5184} X_{\ell_1 \ell_2 \ell_3 \ell_4} X_{i \ell_5 \ell_6 \ell_7}
\Gamma^{\ell_1 \ell_2 \ell_3 \ell_4 \ell_5 \ell_6 \ell_7}
\nn
&&-{1 \over 576} X_{mn \ell_1 \ell_2} X^{mn}{}_{\ell_3 \ell_4} \Gamma_i{}^{\ell_1 \ell_2 \ell_3 \ell_4}
+{1 \over 864} X_{\ell_1 \ell_2 \ell_3}{}^m X_{i \ell_4 \ell_5 m} \Gamma^{\ell_1 \ell_2 \ell_3 \ell_4 \ell_5}
+{1 \over 144} X^{mn}{}_{\ell_1 \ell_2} X_{i \ell_3 mn} \Gamma^{\ell_1 \ell_2 \ell_3}
\nn
&&-{1 \over 432} X_\ell{}^{q_1 q_2 q_3} X_{i q_1 q_2 q_3} \Gamma^\ell
+{1 \over 1728} X_{\ell_1 \ell_2 \ell_3 \ell_4} X^{\ell_1 \ell_2 \ell_3 \ell_4} \Gamma_i
+{1 \over 72} Y_{i \ell_1} Y_{\ell_2 \ell_3} \Gamma^{\ell_1 \ell_2 \ell_3}
+{1 \over 18} Y_{qm} Y_i{}^m \Gamma^q
\nn
&&-{1 \over 108} Y_i{}^m X_{m \ell_1 \ell_2 \ell_3} \Gamma^{\ell_1 \ell_2 \ell_3}
+{1 \over 1728} \Gamma_i{}^{\ell_1 \ell_2 \ell_3 \ell_4 \ell_5 \ell_6}
Y_{\ell_1 \ell_2} X_{\ell_3 \ell_4 \ell_5 \ell_6}
-{1 \over 144} Y^{mn} X_{mn \ell_1 \ell_2} \Gamma_i{}^{\ell_1 \ell_2}
\nn
&&-{1 \over 288} Y_{\ell_1 \ell_2} X_{i \ell_3 \ell_4 \ell_5} \Gamma^{\ell_1 \ell_2 \ell_3 \ell_4 \ell_5}
+{1 \over 48} Y^{mn} X_{imnq} \Gamma^q
+{1 \over 432} X_{\ell_1 \ell_2 \ell_3}{}^m Y_{\ell_4 m} \Gamma_i{}^{\ell_1 \ell_2 \ell_3 \ell_4} \bigg)
(1+e_{1234})=0~.
\nn
\eea
Next consider ({\ref{sp1}}).  Note that
\bea
{1 \over 2} \Gamma^j (\hn_j \hn_i - \hn_i \hn_j) (1+e_{1234}) = {1 \over 4} {\tilde{R}}_{ij} \Gamma^j (1+e_{1234})~,
\eea
where ${\tilde{R}}_{ij}$ denotes the Ricci tensor of ${\cal{S}}$.
Hence, on using ({\ref{sp1}}) to evaluate the LHS of the above expression in terms of the
fluxes $h, X, Y$ and their covariant derivatives, we obtain the following condition
\bea
\label{redx3}
&&\bigg(-{1 \over 4} {\tilde{R}}_{ij} \Gamma^j +{1 \over 8} (\hn_j h_i-\hn_i h_j) \Gamma^j
+{1 \over 72} \Gamma_i{}^{\ell_1 \ell_2 \ell_3} (dY)_{\ell_1 \ell_2 \ell_3}
-{1 \over 12} (dY)_{i \ell_1 \ell_2} \Gamma^{\ell_1 \ell_2}
\nn
&&+{1 \over 12} \hn_i Y_{\ell_1 \ell_2} \Gamma^{\ell_1 \ell_2} +{1 \over 288}
\hn_i X_{\ell_1 \ell_2 \ell_3 \ell_4} \Gamma^{\ell_1 \ell_2 \ell_3 \ell_4}
-{1 \over 48} \Gamma_i{}^{\ell_1 \ell_2 \ell_3} h_{\ell_1} Y_{\ell_2 \ell_3}
-{1 \over 144} \Gamma_i{}^{\ell_1 \ell_2 \ell_3} (i_h X)_{\ell_1 \ell_2 \ell_3}
\nn
&&+{1 \over 8} h_{[i} Y_{\ell_1 \ell_2]} \Gamma^{\ell_1 \ell_2}
+{1 \over 24} (i_h X)_{i \ell_1 \ell_2} \Gamma^{\ell_1 \ell_2}
\nn
&&+{1 \over 41472} \Gamma_i{}^{\ell_1 \ell_2 \ell_3 \ell_4 \ell_5 \ell_6 \ell_7 \ell_8}
X_{\ell_1 \ell_2 \ell_3 \ell_4} X_{\ell_5 \ell_6 \ell_7 \ell_8}
-{1 \over 5184} X_{i \ell_1 \ell_2 \ell_3} X_{\ell_4 \ell_5 \ell_6 \ell_7}
\Gamma^{\ell_1 \ell_2 \ell_3 \ell_4 \ell_5 \ell_6 \ell_7}
\nn
&&-{1 \over 576} \Gamma_i{}^{\ell_1 \ell_2 \ell_3 \ell_4} X^{mn}{}_{\ell_1 \ell_2}
X_{mn \ell_3 \ell_4}
+{1 \over 144} X_{i \ell_1 mn} X^{mn}{}_{\ell_2 \ell_3} \Gamma^{\ell_1 \ell_2 \ell_3}
\nn
&&-{1 \over 864} X_{\ell_1 \ell_2 \ell_3 \ell_4} X^{\ell_1 \ell_2 \ell_3 \ell_4} \Gamma_i
+{1 \over 864} X_{i \ell_1 \ell_2}{}^m X_{\ell_3 \ell_4 \ell_5 m} \Gamma^{\ell_1 \ell_2 \ell_3 \ell_4 \ell_5}
+{1 \over 54} X_{i mnq} X_\ell{}^{mnq} \Gamma^\ell
\nn
&&+{1 \over 1728} \Gamma_i{}^{\ell_1 \ell_2 \ell_3 \ell_4 \ell_5 \ell_6} Y_{\ell_1 \ell_2} X_{\ell_3 \ell_4 \ell_5 \ell_6}
-{1 \over 288} Y_{\ell_1 \ell_2} X_{i \ell_3 \ell_4 \ell_5} \Gamma^{\ell_1 \ell_2 \ell_3 \ell_4 \ell_5}
-{1 \over 432} Y_{m \ell_1} X^m{}_{\ell_2 \ell_3 \ell_4} \Gamma_i{}^{\ell_1 \ell_2 \ell_3 \ell_4}
\nn
&&+{1 \over 108} Y_{mi} X^m{}_{\ell_1 \ell_2 \ell_3} \Gamma^{\ell_1 \ell_2 \ell_3}
-{1 \over 144} Y^{mn} X_{mn \ell_1 \ell_2} \Gamma_i{}^{\ell_1 \ell_2}
+{1 \over 48} Y^{mn} X_{mni\ell} \Gamma^\ell
\nn
&&+{1 \over 72} Y_{i \ell_1} Y_{\ell_2 \ell_3} \Gamma^{\ell_1 \ell_2 \ell_3}
+{1 \over 48} Y^{\ell_1 \ell_2} Y_{\ell_1 \ell_2} \Gamma_i
-{5 \over 72} Y_{im} Y_\ell{}^m \Gamma^\ell \bigg) (1+e_{1234}) =0~.
\nn
\eea
We remark that in order to obtain ({\ref{redx3}}) we have made use of the 4-form field equations
({\ref{geq1}}) and ({\ref{geq2}}) to eliminate divergence terms in $Y$ and $X$ in favour of terms quadratic in the fluxes $h, Y, X$.
Furthermore, the terms involving Hodge duals on the RHS of ({\ref{geq1}}) and ({\ref{geq2}}) have
further been rewritten using the duality between $\Gamma^{i_i \dots i_k}$ and
$\Gamma^{j_1 \dots j_{9-k}}$ when acting on the spinor $1+e_{1234}$.
The closure condition $dX=0$ has also been used in order to simplify
some covariant derivatives acting on $X$.

It remains to compare ({\ref{redx1}}) and ({\ref{redx3}}). Observe that the terms linear in $\hn X$, $\hn Y$,
and the quadratic terms of type $hX$, $hY$, $XY$ match. For the remaining terms in
({\ref{redx1}}) and ({\ref{redx3}}) there is a matching for all orders of Gamma matrices, with the
exception of the 1-Gamma terms. In particular, on taking subtracting ({\ref{redx1}}) from ({\ref{redx3}})
one finds
\bea
\label{redx4}
-{1 \over 4} \bigg( {\tilde{R}}_{ij} + \hn_{(i} h_{j)} -{1 \over 2} h_i h_j
+{1 \over 144} X_{\ell_i \ell_2 \ell_3 \ell_4}X^{\ell_1 \ell_2 \ell_3 \ell_4} g_{ij}
-{1 \over 12} X_{j \ell_1 \ell_2 \ell_3} X_i{}^{\ell_1 \ell_2 \ell_3}
\nn
-{1 \over 12} Y^{\ell_1 \ell_2} Y_{\ell_1 \ell_2} g_{ij}
+{1 \over 2} Y_{im} Y_j{}^m \bigg) \Gamma^j (1+e_{1234}) =0~.
\eea
However, this vanishes as a consequence of ({\ref{ein1}}).

Therefore, we have established that ({\ref{sp2}}) is implied
by ({\ref{sp1}}), ({\ref{thetdef}}), together with  ({\ref{clos}}), the gauge field equations
({\ref{geq1}}) and ({\ref{geq2}}), and the components of the Einstein equations on ${\cal{S}}$.

  \appendix{A Lichnerowicz Identity}

Before we proceed to prove the near horizon  Lichnerowicz identity of section 4, we shall elaborate on $Spin(9)$ spinors. We begin with a realization of Majorana
$Spin(10,1)$  spinors  described
 in appendix A.2 of \cite{system11}. In this, the Dirac spinors are identified with $\Lambda^*(\bC^5)$. The Majorana spinors span a real 32-dimensional
 subspace after an appropriate reality condition is imposed. The Dirac spinors of $Spin(9)$ are identified with the subspace $\Lambda^*(\bC^4)\subset \Lambda^*(\bC^5)$. In particular,
  if $\bC^5=\bC<e_1,\dots e_4,e_5>$, then $\bC^4=\bC<e_1,\dots e_4>$. The Majorana spinors of $Spin(9)$ are those of $Spin(10,1)$ restricted on  $\Lambda^*(\bC^4)$.
 From this, it is straightforward to identify the gamma matrices of $Spin(9)$ from those of $Spin(10,1)$ which have been stated explicitly in \cite{system11}.

 The $Spin(9)$ invariant inner product $\langle \cdot , \cdot \rangle$ is simply the standard Hermitian inner product on $\Lambda^*(\bC^4)$. With respect to this,
 the skew-symmetric products $\Gamma^{[k]}$ of $k$ $Spin(9)$ gamma matrices are Hermitian for $k=0\, {\rm mod}\, 4$ and $k=1\, {\rm mod}\, 4$ while they are anti-Hermitian
 for $k=2\, {\rm mod}\, 4$ and $k=3\, {\rm mod}\, 4$. Using this, we have that
\bea
\Psi_i^\dagger &=& -{1 \over 4} h_i -{1 \over 288} \Gamma_i{}^{\ell_1 \ell_2 \ell_3 \ell_4}
X_{\ell_1 \ell_2 \ell_3 \ell_4} -{1 \over 36} X_{i \ell_1 \ell_2 \ell_3} \Gamma^{\ell_1 \ell_2 \ell_3}
-{1 \over 24} \Gamma_i{}^{\ell_1 \ell_2} Y_{\ell_1 \ell_2} -{1 \over 6} Y_{ij} \Gamma^j~,
\nonumber \\
\Psi^\dagger &=& -{1 \over 4} h_\ell \Gamma^\ell +{1 \over 96} X_{\ell_1 \ell_2 \ell_3 \ell_4}
\Gamma^{\ell_1 \ell_2 \ell_3 \ell_4} -{1 \over 8} Y_{\ell_1 \ell_2} \Gamma^{\ell_1 \ell_2}~,
\eea
where $\dagger$ is the adjoint with respect to the  $Spin(9)$-invariant inner product
$\langle \ , \ \rangle$.

Next let us turn to the computation of the RHS of  ({\ref{lichaux}}).
The term involving $\Psi^{i \dagger} \Psi_i - \Psi^\dagger \Psi$  can be expanded
out directly in terms quadratic in the fluxes $h, Y, X$. In
particular
\bea
\Psi^{i \dagger} \Psi_i &=& {1 \over 16} h^2 +{1 \over 576}
h_{\ell_1} X_{\ell_2 \ell_3 \ell_4 \ell_5} \Gamma^{\ell_1 \ell_2 \ell_3 \ell_4 \ell_5}
+{1 \over 12} (i_h Y)_\ell \Gamma^\ell
\nonumber \\
&-&{5 \over 27648} X_{\ell_1 \ell_2 \ell_3 \ell_4} X_{\ell_5 \ell_6 \ell_7 \ell_8}
\Gamma^{\ell_1 \ell_2 \ell_3 \ell_4 \ell_5 \ell_6 \ell_7 \ell_8}
-{1 \over 384} X^{mn}{}_{\ell_1 \ell_2} X_{mn \ell_3 \ell_4}
\Gamma^{\ell_1 \ell_2 \ell_3 \ell_4}
\nonumber \\
&+&{7 \over 1152} X_{\ell_1 \ell_2 \ell_3 \ell_4} X^{\ell_1 \ell_2 \ell_3 \ell_4}
+{1 \over 192} Y_{\ell_1 \ell_2} Y_{\ell_3 \ell_4} \Gamma^{\ell_1 \ell_2 \ell_3 \ell_4}
+{5 \over 96} Y^{mn} Y_{mn} \ .
\eea

The term in ({\ref{lichaux}}) involving $\Gamma^{ij} \hn_i \hn_j$  can be rewritten using
\bea
\Gamma^{ij} \hn_i \hn_j \phi = -{1 \over 4}\tilde R \phi~,
\eea
where $\tilde R$ is the Ricci scalar of ${\cal{S}}$.  From the Einstein field equation ({\ref{ein1}}), one has
\bea
\tilde R = - \hn^i h_i +{1 \over 2} h^2 +{1 \over 4} Y_{mn} Y^{mn} +{1 \over 48} X_{\ell_1 \ell_2 \ell_3
\ell_4} X^{\ell_1 \ell_2 \ell_3 \ell_4}~.
\eea
It follows that
\bea
{\rm Re \ } \bigg( \int_{\cal{S}} \langle \phi , \Gamma^{ij} \hn_i \hn_j \phi \rangle \bigg)
= \int_{\cal{S}} \langle \phi , \big(-{1 \over 8} h^2 -{1 \over 16} Y_{mn} Y^{mn}
-{1 \over 192} X_{\ell_1 \ell_2 \ell_3 \ell_4} X^{\ell_1 \ell_2 \ell_3 \ell_4} \big) \phi \rangle~,
\eea
where we remark that the contribution to the integral obtained from the $\hn^i h_i$ term is a
total derivative because $ \langle \phi ,\phi \rangle={\rm const}$, and hence its integral vanishes.
Alternatively, this term vanishes if one assumes that $\hn^i h_i=0$.

To proceed with the  evaluation of ({\ref{lichaux}}) observe that
\bea
\label{appaux1}
 \big( \Psi^{i \dagger} - \Psi^i \big) \hn_i \phi
- \big( \Psi^\dagger - \Psi \big) \Gamma^i \hn_i \phi &=&
\big( -{1 \over 72} X_{\ell_1 \ell_2 \ell_3 \ell_4} \Gamma^{i \ell_1 \ell_2 \ell_3 \ell_4}
-{1 \over 6} Y^i{}_\ell \Gamma^\ell \big) \hn_i \phi
\nonumber \\
&+& \big( {1 \over 72} X_{\ell_1 \ell_2 \ell_3 \ell_4} \Gamma^{\ell_1 \ell_2 \ell_3 \ell_4}
+{1 \over 6} Y_{\ell_1 \ell_2} \Gamma^{\ell_1 \ell_2} \big) \Gamma^i \hn_i \phi~.
\nonumber \\
\eea
Using the fact that the Clifford algebra element of first term in the RHS of the above equation is self-adjoint, the Bianchi identity $dX=0$ and upon integrating by parts, one finds that
\bea
{\rm Re \ } \bigg( \int_{\cal{S}} \langle \phi , \big( \Psi^{i \dagger} - \Psi^i \big) \hn_i \phi
- \big( \Psi^\dagger - \Psi \big) \Gamma^i \hn_i \phi \rangle \bigg)
= \int_{\cal{S}} \langle \phi, {1 \over 12} (\hn^i Y_{i \ell}) \Gamma^\ell \phi \rangle
\nonumber \\
+ {\rm Re \ } \bigg( \int_{\cal{S}} \langle \phi , \big( {1 \over 72} X_{\ell_1 \ell_2 \ell_3 \ell_4}
\Gamma^{\ell_1 \ell_2 \ell_3 \ell_4} +{1 \over 6} Y_{\ell_1 \ell_2} \Gamma^{\ell_1 \ell_2}
\big) \Gamma^i \hn_i \phi \rangle \bigg) \ .
\eea
The term involving $\hn^i Y_{i \ell}$ is then further rewritten as a term quadratic in $X$ using the field equation
({\ref{geq2}}). Next, we rewrite the second line in terms of the Dirac operator $\Gamma^i \hn_i \phi + \Psi \phi$,
with a compensating term $-\Psi \phi$ which gives a term quadratic in the fluxes $h, X, Y$, and
which can be expanded out straightforwardly.

Next, we find that
\bea
\label{appaux2}
\big( \Gamma^i \Psi - \Psi \Gamma^i \big) \hn_i \phi &=& \big( -{1 \over 2} h^i +{1 \over 48} \Gamma^{i \ell_1 \ell_2 \ell_3 \ell_4} X_{\ell_1 \ell_2 \ell_3 \ell_4} +{1 \over 2} Y^i{}_\ell \Gamma^\ell \big) \hn_i \phi
\nonumber \\
&+& \big( {1 \over 2} h_\ell \Gamma^\ell -{1 \over 48} X_{\ell_1 \ell_2 \ell_3 \ell_4}
\Gamma^{\ell_1 \ell_2 \ell_3 \ell_4} \big) \Gamma^i \hn_i \phi~.
\eea
Similarly, using the self-duality of the Clifford element in the first term in the RHS of the above equation, $dX=0$ and the condition $ \langle \phi ,\phi \rangle={\rm const}$, and upon integrating by parts, one finds
\bea
{\rm Re \ } \bigg(\int_{\cal{S}} \langle \phi , \big( \Gamma^i \Psi - \Psi \Gamma^i \big) \hn_i \phi \rangle \bigg)
&=& \int_{\cal{S}} \langle \phi , -{1 \over 4} (\hn^i Y_{i \ell}) \Gamma^\ell \phi \rangle
\nonumber \\
&+& {\rm Re \ } \bigg( \int_{\cal{S}} \langle \phi , \big( {1 \over 2} h_\ell \Gamma^\ell
-{1 \over 48} X_{\ell_1 \ell_2 \ell_3 \ell_4} \Gamma^{\ell_1 \ell_2 \ell_3 \ell_4} \big)
\Gamma^i \hn_i \phi \rangle \bigg) \ .
\nonumber \\
\eea
The term involving $\hn^i Y_{i \ell}$ is then further rewritten as a term quadratic in $X$ using
({\ref{geq2}}). The second line is also
further rewritten  in terms of the Dirac operator $\Gamma^i \hn_i \phi + \Psi \phi$,
with a compensating term involving $-\Psi \phi$ which gives a term quadratic in the fluxes $h, X, Y$.

Next note that
\bea
(\Gamma^i \hn_i \Psi - \hn^i \Psi_i) \phi
&=& \bigg( -{1 \over 8} dh_{ij} \Gamma^{ij} +{1 \over 72} \hn^i X_{i \ell_1 \ell_2 \ell_3}
\Gamma^{\ell_1 \ell_2 \ell_3}
\nonumber \\
&+&{1 \over 12} \hn_{\ell_1} Y_{\ell_2 \ell_3}
\Gamma^{\ell_1 \ell_2 \ell_3} +{5 \over 12} \hn^i Y_{i \ell} \Gamma^\ell \bigg) \phi
\nonumber \\
\eea
and so
\bea
{\rm Re \ } \bigg( \int_{\cal{S}} \langle \phi , (\Gamma^i \hn_i \Psi - \hn^i \Psi_i) \phi\rangle \bigg)
= \int_{\cal{S}} \langle \phi , {5 \over 12} \hn^i Y_{i \ell} \Gamma^\ell  \phi \rangle \ ,
\eea
as the rest of the terms are anti-self-adjoint, and hence imaginary, and so they vanish.
The term involving $\hn^i Y_{i \ell}$ is then again rewritten as a term quadratic in $X$ using
({\ref{geq2}})

\appendix{Solution of KSEs on $M^8$}

An alternative way to solve the KSEs (\ref{gravr8}) and (\ref{ux2}) is to choose representatives for $\phi^\pm$ as
\bea
\phi^+=f (1+e_{1234})~,~~~\phi^-=g(e_1+e_{234})
\eea
where $f,g$ are real functions which may vanish at some points on $M^8$. This choice can always
be made using the local $Spin(8)$ covariance of the equations.

First consider the KSE (\ref{gravr8}) with the Levi-Civita connection acting on $\phi^-$. This gives
\bea
\partial_A g+{g\over2} \Omega_{A,1\bar 1}-{g\over 2} \Omega_{A, p}{}^p + {\sqrt 2 m \over 4} f \delta_{A\bar 1}
+{\sqrt 2\over12}
f X_{Apqr} \e^{pqr}+{\sqrt 2\over 4} f X_{A\bar 1p}{}^p =0~,
\la{parm8}
\eea
and
\bea
{g\over 2} \Omega_{A, rs} \e^{rs}{}_{\bar p}+ g \Omega_{A, 1\bar p}+ {\sqrt 2m\over 4} f \delta_{A\bar p}
-{\sqrt 2\over 4} f  X_{A1 rs} \e^{rs}{}_{\bar p}+ {\sqrt 2\over 4} f X_{A\bar p\g}{}^\g=0~,
\eea
where in this section the indices $A=(\a, \bar \a)$ with $\a=(1, p)$ and $\bar\a=(\bar 1, \bar p)$, and $p=2,3,4$.
The conditions which arise from the  other KSEs with the Levi-Civita connection acting on $\phi^+$
are derived from the above after exchanging $1\leftrightarrow  -\bar 1$, $g\rightarrow f$ and $f\rightarrow-g$  and $m\rightarrow -m$. The equations decompose under the $SU(3)$ subgroup
of $G_2$ which is the isotropy group of the spinors $\phi^\pm$ at non-vanishing points.

Similarly the solution of the (\ref{ux2}) KSE on the spinor $\phi^+$ is
\bea
{1\over 18} X_{\a_1\a_2\a_3\a_4}\e^{\a_1\a_2\a_3\a_4} +{1\over 6} X_\a{}^\a{}_\b{}^\b+m&=&0
\nonumber \\
{1\over 2} X_\a{}^\a{}_{\b_1\b_2} \e^{\b_1\b_2}{}_{\bar\g_1\bar\g_2}+X_\a{}^\a{}_{\bar\g_1\bar\g_2}&=&0 \ .
\la{parm8b}
\eea
The conditions on the $X^-$ can be derived by exchanging $1\leftrightarrow -\bar 1$ and $m\rightarrow -m$.

Next observe that the parallel transport equations, after using (\ref{parm8b}), give
\bea
\partial_A g+ Y_A f=0~,~~~\partial_A f-Y_Ag=0~,~~~Y_A=-{\sqrt 2 m\over 2 } \mathrm {Re} [\delta_{A\bar 1}]~.
\la{parrap}
\eea
Thus $f, g$ depend on one direction in $M^8$, and $f^2+g^2$ is constant which we set to $1$. So although their vanishing locus may not be isolated
their measure is zero because otherwise both must vanish at the same point which is not allowed.
As a consequence, although $g, f$ multiply the conditions on the fluxes and so they do not impose a
condition at the vanishing locus, the fluxes can be determined everywhere in terms of the geometry
as a consequence of continuity.  Furthermore, observe that
\bea
d(gf Y)=0
\eea
as an integrability condition of (\ref{parrap}), which agrees with (\ref{closv1}) as $fg Y$
is proportional to $\xi$.

To solve the   linear system  (\ref{parm8}) and (\ref{parm8b}), set $g=\sin\theta$ and $f=\cos\theta$
for some angle $\theta$ which depends on spacetime points.  Then the fluxes can be expressed
in terms of the geometry as
\bea
X_{ 1\bar p q}{}^q&=&-{2\over \sqrt2} \tan\theta\, \Omega_{1,mn} \e^{mn}{}_{\bar p}-{4\over \sqrt2}
\tan\theta\, \Omega_{1,1\bar p}~,~~~
\cr
X_{ \bar1\bar p q}{}^q&=&{2\over \sqrt2} \cot\theta\, \Omega_{\bar1,mn} \e^{mn}{}_{\bar p}-{4\over \sqrt2}
\cot\theta\, \Omega_{\bar1,\bar1\bar p}~,
\cr
X_{rsq}{}^q&=&{1\over2}\e^{\bar p}{}_{rs} (X_{ 1\bar p q}{}^q+X_{ \bar1\bar p q}{}^q)~,~~~
X_{rs1\bar 1}=-{1\over2} \e^{\bar p}{}_{rs} (X_{ 1\bar p q}{}^q-X_{ \bar1\bar p q}{}^q)~,
\cr
X_{q\bar p r}{}^r&=&- D^-\delta_{q\bar p}-\sqrt 2\big[{\tan\theta-\cot\theta\over2} \Omega_{q,rs} \e^{rs}{}_{\bar p}
+\tan\theta\, \Omega_{q,1\bar p}+\cot\theta\, \Omega_{q,\bar1\bar p}\big]~,
\cr
X_{p\bar q 1\bar 1}&=&( D^+-m)\delta_{p\bar q}
-\sqrt2 \big[{\tan\theta+\cot\theta\over2} \Omega_{q,rs} \e^{rs}{}_{\bar p}
+\tan\theta\, \Omega_{q,1\bar p}-\cot\theta\, \Omega_{q,\bar1\bar p}\big]~,
\cr
X_{1rst}\e^{rst}&=&{3\over2} D^--{3\over2} D^+~,~~~ X_{\bar1rst}\e^{rst}=-{3\over2} D^--{3\over2} D^+~,
\cr
{\sqrt 2\over 4} X_{\bar q 1rs} \e^{rs}{}_{\bar p}&=&{1\over 2}\tan\theta \, \Omega_{(\bar q,| rs|} \e^{rs}{}_{\bar p)}+ \tan \theta \Omega_{(\bar q,|1|\bar p)}+{\sqrt 2\over 4} \e^m{}_{\bar q\bar p}
X_{m1\ell}{}^\ell~,
\cr
{\sqrt 2\over 4} X_{\bar q \bar1rs} \e^{rs}{}_{\bar p}&=&{1\over 2}\cot\theta \, \Omega_{(\bar q,| rs|} \e^{rs}{}_{\bar p)}- \cot \theta \Omega_{(\bar q,|\bar1|\bar p)}+{\sqrt 2\over 4} \e^m{}_{\bar q\bar p}
X_{m\bar1\ell}{}^\ell~,
\eea
where
\bea
D^\pm=\sqrt2 \big[{\tan\theta\pm \cot\theta\over2}+\tan\theta \Omega_{q,1}{}^q\mp \cot\theta \Omega_{q,\bar1}{}^q\big]~.
\eea
The conditions on the geometry of $M^8$ implied by the linear system  are
\bea
&&(\Omega_{1,rs}+\Omega_{\bar 1, rs}) \e^{rs}{}_{\bar p}+ 2 (\Omega_{1,1\bar p}+\Omega_{\bar 1,1\bar p})=0~,~~~
(\Omega_{1,rs}+\Omega_{\bar 1, rs}) \e^{rs}{}_{\bar p}- 2 (\Omega_{1,\bar 1\bar p}+\Omega_{\bar 1,\bar1\bar p})=0~,
\cr
&& \Omega_{[\bar q,|rs|} \e^{rs}{}_{\bar p]}+2\Omega_{[\bar q, |1|\bar p]}=0~,~~~
\Omega_{[\bar q,|rs|} \e^{rs}{}_{\bar p]}-2\Omega_{[\bar q, |\bar 1|\bar p]}=0~,
\cr
&&\Omega_{q,1\bar1}-\Omega_{q,p}{}^p-2\Omega_{\bar 1,\bar 1 q}-\Omega_{\bar 1, \bar r\bar s} \e^{\bar r\bar s}{}_q=0~,~~~\Omega_{q,1\bar1}+\Omega_{q,p}{}^p+2\Omega_{1,1 q}-\Omega_{ 1, \bar r\bar s} \e^{\bar r\bar s}{}_q=0~,
\cr
&&\Omega_{q,rs} \e^{rs}{}_{\bar p}-\Omega_{\bar p,\bar r\bar s} \e^{\bar r\bar s}{}_{ q}
+2 \Omega_{q, 1\bar p}-2  \Omega_{\bar p, \bar 1 q}=0~,~~\Omega_{q,rs} \e^{rs}{}_{\bar p}-\Omega_{\bar p,\bar r\bar s} \e^{\bar r\bar s}{}_{ q}
-2 \Omega_{q, \bar1\bar p}+2  \Omega_{\bar p,  1 q}=0
\cr
&&-\Omega_{1,p}{}^p+\Omega_{r,st} \e^{rst}
-{1\over 2\sqrt2} m (\cot\theta+\tan\theta)=0~,~~
\cr
&&-\Omega_{1,1\bar1} -2 \Omega_{q,\bar1}{}^q
+{1\over2\sqrt2} m
(\cot\theta-\tan\theta)=0~.
\eea
These geometric conditions must also be supplemented with $d(e^1+e^{\bar 1})=0$ which follows
from (\ref{parrap}).
The above conditions can be re-expressed in terms of $G_2$ representations but their  form
above in terms of $SU(3)$ representations suffices.

\end{document}